\documentclass[a4paper,journal]{IEEEtran}

\usepackage{graphicx}
\usepackage{amssymb}
\usepackage{amsmath}
\usepackage{epstopdf}
\usepackage{subfig}
\usepackage{pstricks}
\usepackage{tabularx,booktabs}
\usepackage{mathtools}
\usepackage{algorithm}
\usepackage{algorithmic}
\usepackage{rotating}
\usepackage{bm}
\usepackage{framed}
\colorlet{shadecolor}{yellow!50}
\newcolumntype{L}[1]{>{\raggedright\let\newline\\\arraybackslash}m{#1}}
\newcolumntype{C}[1]{>{\centering\let\newline\\\arraybackslash}m{#1}}
\newcolumntype{R}[1]{>{\raggedleft\let\newline\\\arraybackslash}m{#1}}

\def \CoS {\mathbb{C}} 
\def \ReS {\mathbb{R}} 
\def \H {^H}

\def \l {\ell}

\def \m {\mathbf m}

\def \x {\mathbf x}
\def \X {\mathbf X}
\def \xh {\mathbf {\hat x} }
\def \Xh {\mathbf {\hat X} }
\def \xs {\mathbf {x}}
\def \XS {\mathbf {X}}

\def \Xs {\mathbf {\bar X}}
\def \Xsh {\mathbf {\hat {\bar X}}}
\def \t {\mathbf t}

\def \v {\mathbf v}
\def \y {\mathbf y}

\def \z {\mathbf z}
\def \Z {\mathbf Z}

\def \dcf {\delta_{{\text {CF}}}}

\DeclareMathOperator{\Tr}{Tr}
\DeclareMathOperator*{\argmin}{arg\,min}

\DeclareMathOperator{\R}{R}
\DeclareMathOperator{\phase}{Phase}

\def \AmpCal {\textit {\textbf {A-Cal}}}
\def \PhaseCalJ {\textit {\textbf {P-Cal}}}

\def \PhaseCalS {\textit {\textbf {P-Cal-Scalable}}}
\def \CompCalJ {\textit {\textbf {C-Cal}}}
\def \CompCalS {\textit {\textbf {C-Cal-Scalable}}}

\begin{document}

\title{Convex Optimization Approaches for Blind Sensor Calibration using Sparsity}

\author{%
\IEEEauthorblockN{
\c{C}a\u{g}da\c{s}~Bilen\IEEEauthorrefmark{1}, 
Gilles Puy\IEEEauthorrefmark{2}\IEEEauthorrefmark{1}, 
R\'emi Gribonval\IEEEauthorrefmark{1} 
and Laurent Daudet\IEEEauthorrefmark{3}}
\\
\IEEEauthorblockA{\IEEEauthorrefmark{1} 
INRIA, Centre Inria Rennes - Bretagne Atlantique, 35042 Rennes Cedex, France.}
\\
\IEEEauthorblockA{\IEEEauthorrefmark{2}
Inst. of Electrical Eng., Ecole Polytechnique Federale de Lausanne (EPFL) CH-1015 Lausanne, Switzerland}
\\
\IEEEauthorblockA{\IEEEauthorrefmark{3}
Institut Langevin, CNRS UMR 7587, UPMC, Univ. Paris Diderot, ESPCI, 75005 Paris, France \thanks{This work was partly funded by the Agence Nationale de la Recherche (ANR), project ECHANGE (ANR-08-EMER-006) and by the European Research Council, PLEASE project (ERC-StG-2011-277906). LD is on a joint affiliation between Univ. Paris Diderot and Institut Universitaire de France.}}
}

\maketitle

\begin{abstract}
We investigate a compressive sensing framework in which the sensors introduce a distortion to the measurements in the form of unknown gains. We focus on {\em blind} calibration, using measures performed on {\em multiple} unknown (but sparse) signals and formulate the joint recovery of the gains and the sparse signals as a convex optimization problem. We divide this problem in 3 subproblems with different conditions on the gains, specifially (i) gains with different amplitude and the same phase, (ii) gains with the same amplitude and different phase and (iii) gains with different amplitude and phase. In order to solve the first case, we propose an extension to the basis pursuit optimization which can estimate the unknown gains along with the unknown sparse signals. For the second case, we formulate a quadratic approach that eliminates the unknown phase shifts and retrieves the unknown sparse signals. An alternative form of this approach is also formulated to reduce complexity and memory requirements and provide scalability with respect to the number of input signals. Finally for the third case, we propose a formulation that combines the earlier two approaches to solve the problem. The performance of the proposed algorithms is investigated extensively through numerical simulations, which demonstrates that simultaneous signal recovery and calibration is possible with convex methods when sufficiently many (unknown, but sparse) calibrating signals are provided.
\end{abstract}


\section{Introduction}
Compressed sensing is a theoretical and numerical framework to sample sparse signals at lower rates than required by the Nyquist-Shannon theorem \cite{Donoho2006}. More precisely, a $K$-sparse source vector, $\xs \in \CoS^N$, i.e. a source vector with only $K$ non-zero entries, is sampled by a number $M$ ($<N$) of linear measurements 
\begin{align}
\label{eq:calibrated_measurement_model}
y_{i} = \m_i\H \xs, \qquad i = 1,\ldots, M
\end{align}
where $\m_1, \ldots, \m_M \in \CoS^N$ are \emph{known} measurement vectors, and $\mathbf{.}\H$ denotes the conjugate transpose operator. Under certain conditions on the measurement vectors, the signal can be reconstructed accurately by solving the basis pursuit problem, e.g.,
\begin{align}
\label{eq:CalibratedL1}
\xh_{\ell_1} = &\argmin_{\z} \Vert \z \Vert_1\\
\nonumber &\text{subject to}\quad y_i = \m_i\H \z,\quad i = 1,\ldots, M
\end{align}
where $\Vert \mathbf{\cdot} \Vert_1$ denotes the $\ell_1$-norm, which favors the selection of sparse signals among the ones satisfying the measurement constraints. It has been shown that the number of measurements needed for accurate recovery, such that $\xh_{\ell_1} = \xs$, scales only linearly with $K$ \cite{Donoho2006}. Note that the above minimization problem can be modified easily to handle the presence of additive noise on the measurements.

In practice, it is often not possible to know the measurement vectors $\m_1, \ldots, \m_M$ perfectly. In many applications dealing with distributed sensors or radars, the location or intrinsic parameters of the sensors are not known exactly \cite{Chong1996, Balzano2007, Wang2008, Yang2012}, which in turn may result in unknown phase shifts and/or gains at each sensor. Similarly, applications with microphone arrays are shown to require calibration of each microphone to account for the unknown gain and phase shifts introduced at each frequency \cite{Mignot2011}. Unlike additive perturbations in the measurement matrix, compressive sensing reconstruction is not robust to multiplicative perturbation \cite{Candes2006, Herman2010}. Consequently various calibration scenarios for the measurement matrix in compressive sensing have been investigated in the recent years. Perturbations in a parameterized measurement matrix have been estimated along with the signal in \cite{Johnson2013}. Calibration for unknown scaling of the input signal using the generalized approximate message passing (GAMP) algorithm has been considered in \cite{Kamilov2013}. The scenario of calibration for the unknown gains introduced by the sensors has been investigated in \cite{Schulke2013}, also using GAMP. We consider the same scenario in this paper and propose various convex optimization methods to calibrate the measurement system automatically while estimating the unknown sparse input signals.


\begin{figure*}[!t]
\centering
\psscalebox{0.90}{
\psset{unit=0.26cm}
\input{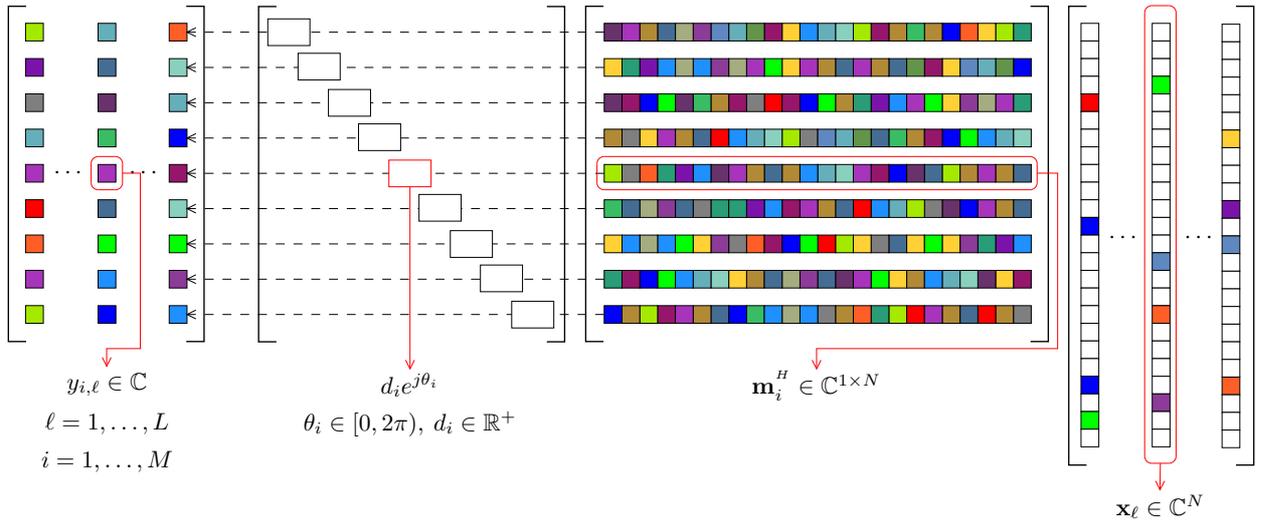}
}
\caption{The system diagram depicting the relationship between the unknown input sparse signals, $\xs_\l$, the sensing vectors, $\m_i$, the unknown sensor gains, $d_i e^{j\theta_i}$, and the measured output signals, $\y_{i,\l}$.}
\label{fig:system}
\end{figure*}

In this paper, we investigate the problem of blindly calibrating unknown complex-valued gains that are common for each set of measurements taken by the same sensor in a compressive sensing system. By ``blindly", we mean that our goal here is to retrieve the unknown gains along with sparse input signals, assuming that multiple input signals are measured with the same sensor system. We start by considering two different special cases of this problem, namely amplitude calibration and phase calibration. In the amplitude calibration problem the phase shifts introduced by the gains are assumed to be known, whereas the phase calibration problem considers the opposite case of known amplitudes of the gains. We propose a convex optimization algorithm to solve amplitude calibration and analyze the performance of the proposed algorithm in Section~\ref{sec:AmpCal}. Similarly, we propose two different algorithms to solve the phase calibration problem in Sections~\ref{sec:PhaseCal} and \ref{sec:PhaseCalSc}. We then consider the general gain calibration problem in Section~\ref{sec:CompCal} and propose two convex algorithms derived by combining the algorithms introduced in the earlier sections. Conclusions and discussions on future work are presented in Section~\ref{sec:Conc}.

\section{Problem Definition}
\label{sec:PDef}

Suppose that the measurement system in \eqref{eq:calibrated_measurement_model} is perturbed by unknown constant complex gains, $d_i e^{j\theta_i}$, at each sensor $i$ and that multiple sparse input signals, ${\xs}_\l \in \CoS^{N}, \quad \l=1\dots L$, are measured through the system such that
\begin{align}
\label{eq:transfunc}
& y_{i,\l} =d_i e^{j\theta_i} \m_i\H {\xs}_\l \qquad i=1,\dots ,M,\; \theta_i \in [0,2\pi),\:d_i >0
\end{align}
and the measurements $y_{i,\l}$ and the measurement vectors $\m_i$ are known. A visual depiction of this measurement system with unknown sensor gains can be seen in Figure~\ref{fig:system}. This system can be characterized by two main conditions on the gains called \textbf{amplitude variability} and \textbf{phase variability}. Amplitude variability is the variance of the amplitudes, $d_i$, of the gains among different sensors whereas the phase variability is the variance of phases, $\theta_i$, of the gains. 

The problem of calibrating the gains in a system described by \eqref{eq:transfunc} arises in a number of practical applications. For instance the sensor networks composed of a large number of sensors may suffer from faulty or uncalibrated sensors with unknown amplification as described in \cite{Balzano2007}. Microphone arrays are also known to have varying amplification depending on the setup and the quality of microphones used \cite{Mignot2011}. In radar imaging applications such as inverse synthetic aperture radar (ISAR) imagery, the measurements suffer from small phase shifts due to small movements of the target object during imaging, resulting in exactly the system of equations shown in \eqref{eq:transfunc} with known $d_i$ \cite{Zhao2014}. Last but not least, the vast number of blind deconvolution problems in various applications can be transformed into a complex valued gain calibration problem as in \eqref{eq:transfunc} with the Fourier transform of the entire set of equations and measurements. Hence the blind calibration for complex valued gains in a sparse inverse problem is highly relevant to a large number of application fields in signal processing.

It has been shown that ignoring the unknown gains during recovery, i.e. treating the decalibration as noise, can significantly reduce recovery performance, especially when there is ambiguity in phase \cite{Herman2010}. Therefore it is essential to employ a reconstruction approach that deals with unknown gains rather than ignoring them.

In a compressed sensing scenario the signal is estimated from incomplete measurements ($M<N$). However in the calibration problem, the overcomplete case ($M>N$) is also of interest since more measurements may be needed to compensate for the increased number of unknowns. We focus on the noiseless case only for the sake of simplicity. Some discussions on the noisy case are provided in Section~\ref{sec:Conc}. We also focus on the complex-valued case only, however all of the derivations and the proposed methods can also be applied to real-valued systems with the phase term $e^{j\theta_i}$ being replaced with $\pm 1$.

In a traditional recovery strategy, one can enforce the sparsity of the input signals while enforcing the measurement constraints in \eqref{eq:transfunc}. However, when dealing with unknown gains, the measurement constraints are non-linear with respect to the unknowns $d_i$ and ${\xs}_\l$. This non-linearity can be tackled using an alternating minimization strategy where one estimates $\xs$ while keeping $d_i$ fixed and vice-versa iteratively \cite{Yang2012}. However, the convergence of this alternating optimization to the global minimum is not guaranteed.

In this paper, we formulate the problem of joint signal recovery and calibration from the measurements $y_{i,\l}$ in \eqref{eq:transfunc} as a convex optimization problem. We shall consider three main scenarios on the sensor characteristics, each of which will be approached and formulated differently:
\begin{enumerate}
\item \textbf{Amplitude Calibration}: small or no phase variability, discussed in Section~\ref{sec:AmpCal}.
\item \textbf{Phase Calibration}: large phase variability but no amplitude variability, discussed in Sections~\ref{sec:PhaseCal} and \ref{sec:PhaseCalSc}.
\item \textbf{Complete Calibration}: large amplitude and phase variability, discussed in Section~\ref{sec:CompCal}.
\end{enumerate}

\section{Amplitude Calibration}
\label{sec:AmpCal}

In this section, we first consider the case where the phase variability is small (or, equivalently, approximately known). Similar to the approach proposed in \cite{Gribonval2012}, it is then possible to overcome the non-linearity in the measurement equation \eqref{eq:transfunc} with respect to $d_i e^{j\theta_i}$ and ${\xs}_\l$ by re-parameterizing it in a linear fashion such that
\begin{align}
\label{eq:transfunc2}
 y_{i,\l}\tau_i = \m_i\H {\xs}_\l \qquad &i=1,\dots, M\;,\; \l=1,\dots, L\\
\nonumber &\tau_i \triangleq \frac{1}{d_i e^{j\theta_i}}
\end{align}
Similar to \eqref{eq:transfunc}, there is a global scale and phase ambiguity between the gains and the input signals in these equations, i.e. all the constraints in \eqref{eq:transfunc2} can also be satisfied with the set of gains $\{\frac{d_i e^{j\theta_i}}{c_0} \}$ and the set of equally sparse signals $\{c_0 \xs_\l\}$ for any non-zero $c_0 \in \CoS\setminus \{0\}$. However unlike \eqref{eq:transfunc}, \eqref{eq:transfunc2} is also satisfied by the trivial solution ($\tau_i=0$, $\xs_\l=0$ for $i=1,\ldots,M$, $\l=1,\ldots,L$). To avoid this problem, we add one more constraint 
\begin{align}
\label{eq:sumconst}
\sum_{i=1}^M \tau_i = c 
\end{align}
in which $c \in \CoS\setminus \{0\}$ is an arbitrary constant. The only drawback of having the constraint in \eqref{eq:sumconst} is that it excludes the solutions where the sum of the inverse of the gains are zero ($\sum_{i} \frac{1}{d_i e^{j \theta_i}} = 0$) from the solution set which is possible for complex-valued gains in theory. However this does not pose a problem since the sum being exactly equal to zero is highly unlikely in practical cases.

We can observe from the set of equations in \eqref{eq:transfunc2} and \eqref{eq:sumconst} that there are $NL+M$ unknowns (the $\xs_\l$ and $\tau_i$) which are constrained with $ML$ measurements and the sum constraint. Assuming these $ML+1$ equations are consistent and complete, a unique solution can be found provided that 
\begin{align}
\label{eq:deltaBound}
\frac{M}{N} \geq \dcf,\quad \dcf \triangleq \frac{NL-1}{NL-N} \cong (1+\frac{1}{L-1})
\end{align}
The bound $\dcf N$ indicates the minimum number of measurements enabling a unique closed form solution for the blind calibration problem regardless of the sparsity of the input signals. Hence when \eqref{eq:deltaBound} is satisfied, blind calibration is simply a matter of solving the linear equations in \eqref{eq:transfunc2} and \eqref{eq:sumconst}. 

When the input signals are sparse and $M<\dcf N$, it is possible to search for the sparsest solution consistent with \eqref{eq:transfunc2} and \eqref{eq:sumconst} and attempt to recover the sparse signals and the gains with the convex optimization approach
\begin{align}
\nonumber \AmpCal\textbf{:} \qquad\qquad\qquad\qquad &\\
\label{eq:opt}
\begin{array}{c}
{\xh}_1,\ldots,{\xh}_L\\
\hat{\tau}_1,\ldots,\hat{\tau}_M
\end{array}
 = \;\argmin_{\substack{\z_1,\ldots,\z_L\\ t_1,\ldots,t_M}} \quad &\sum_{n=1}^L \Vert \z_n \Vert_1 \\
\nonumber \text{subject to}\quad &y_{i,\l} t_i = \m_i\H \z_\l \qquad \begin{array}{l}
\l=1,\ldots,L\\
i=1,\ldots,M
\end{array}\\
\nonumber & \sum_{n=1}^M t_n = c 
\end{align}
for an arbitrary constant $c>0$. The actual gains are recovered after the optimization using $\hat{d}_i e^{j\hat{\theta}_i} = \frac{1}{\hat{\tau}_i}$. 



\subsection{Experimental Results for Amplitude Calibration}
\label{sec:AmpCalExp}
\begin{figure*}[t]
\centering
\subfloat{
\rotatebox{90}{
\begin{minipage}[t]{.01\textwidth}
{ }
\end{minipage}
}}
\subfloat{
\makebox[.23\textwidth][c]{{\footnotesize $\qquad p_c=0$}}
}
\subfloat{
\makebox[.23\textwidth][c]{{\footnotesize $\qquad p_c=1/3$}}
}
\subfloat{
\makebox[.23\textwidth][c]{{\footnotesize $\qquad p_c=2/3$}}
}
\subfloat{
\makebox[.23\textwidth][c]{{\footnotesize $\qquad p_c=1$}}
}
\\
\subfloat{
\rotatebox{90}{
\begin{minipage}[t]{.14\textwidth}
{\begin{center}
\tiny $\quad\: L=5,\:\sigma=0.1$
\end{center}}
\end{minipage}
}}
\subfloat{
\includegraphics[width=.23\textwidth]{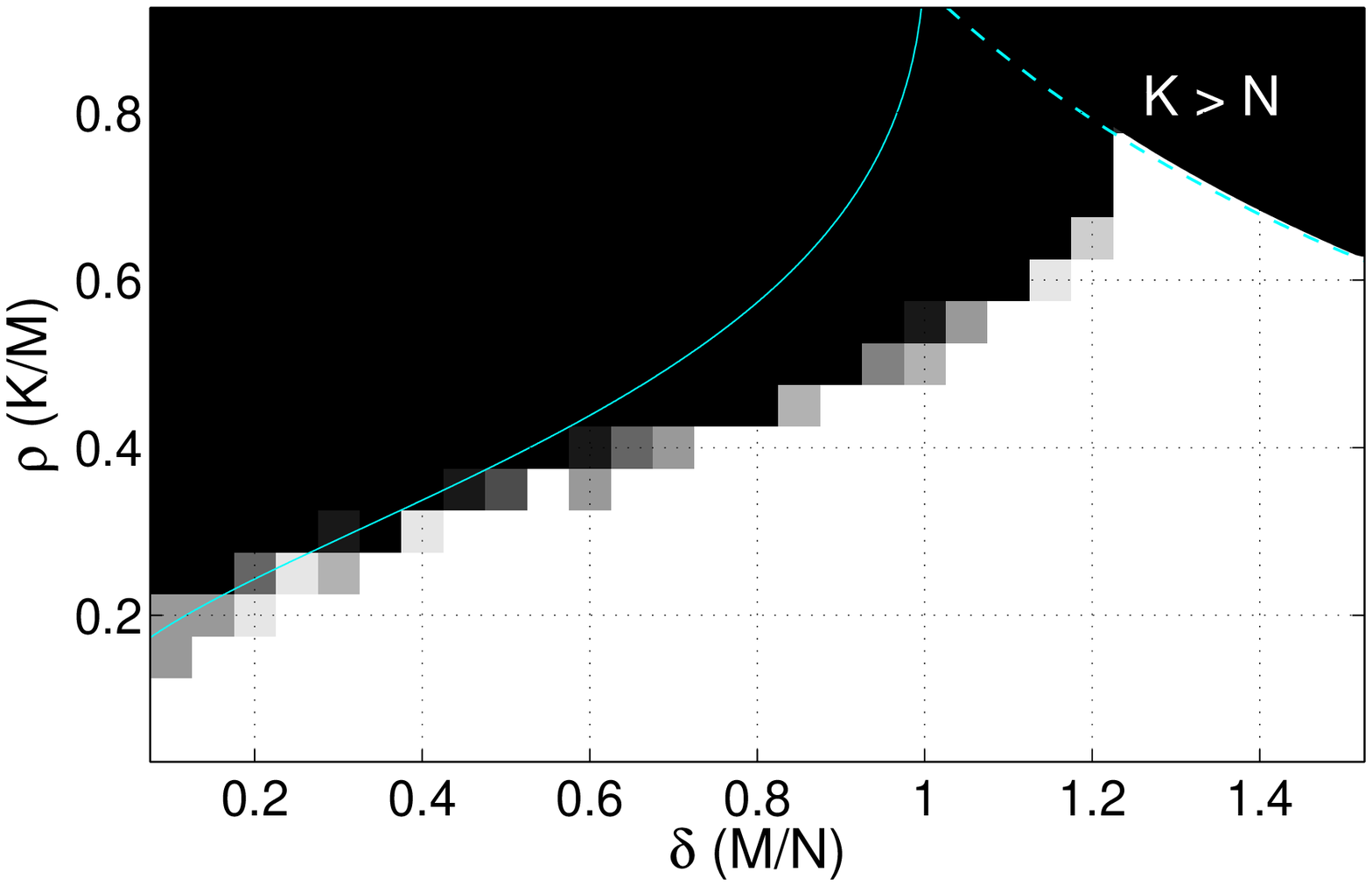}
}
\subfloat{
\includegraphics[width=.23\textwidth]{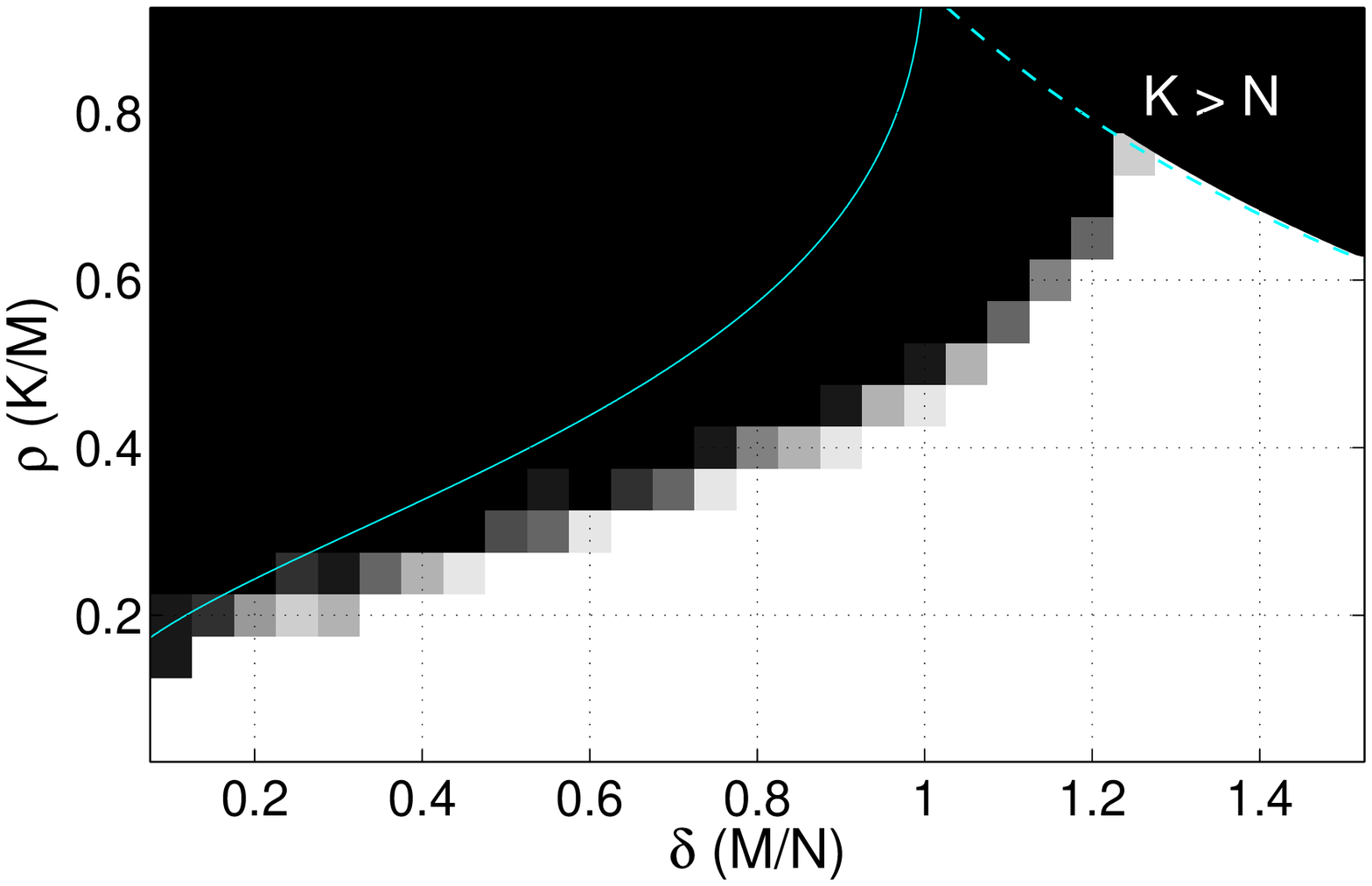}
}
\subfloat{
\includegraphics[width=.23\textwidth]{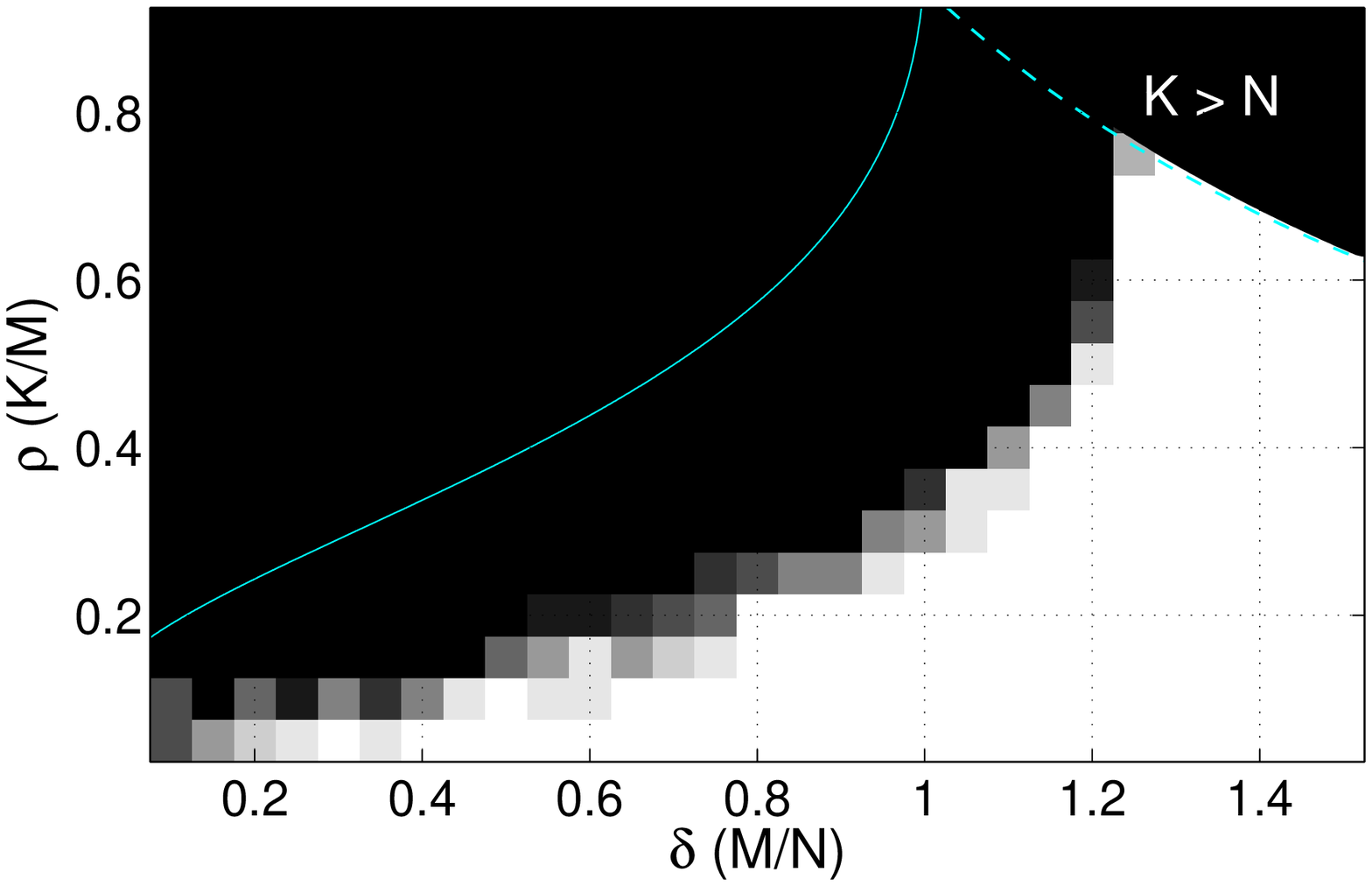}
}
\subfloat{
\includegraphics[width=.23\textwidth]{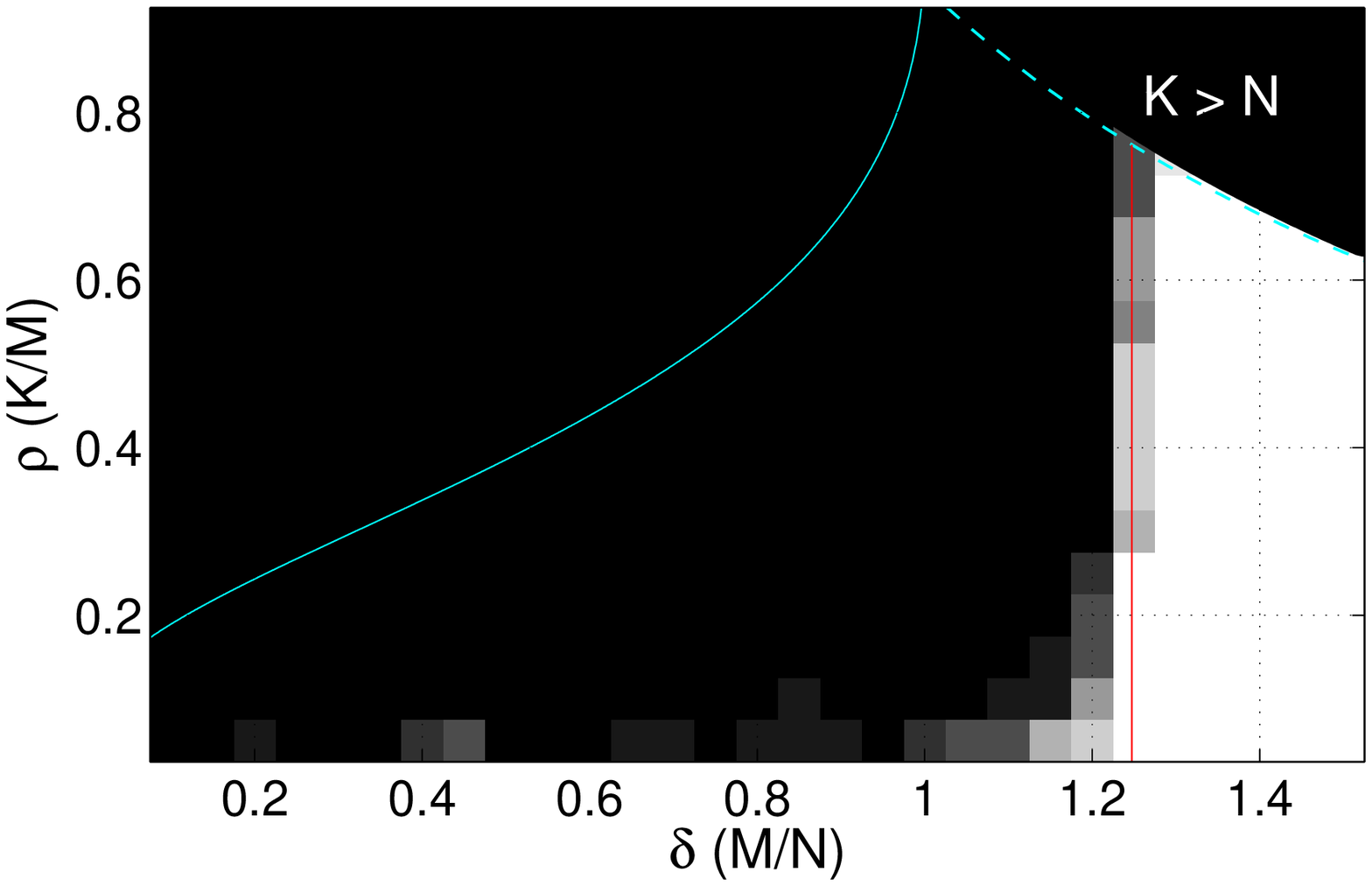}
}
\\
\subfloat{
\rotatebox{90}{{
\begin{minipage}[t]{.14\textwidth}
{\begin{center}
\tiny $\quad\: L=10,\:\sigma=0.3$
\end{center}}
\end{minipage}
}}
}
\subfloat{
\includegraphics[width=.23\textwidth]{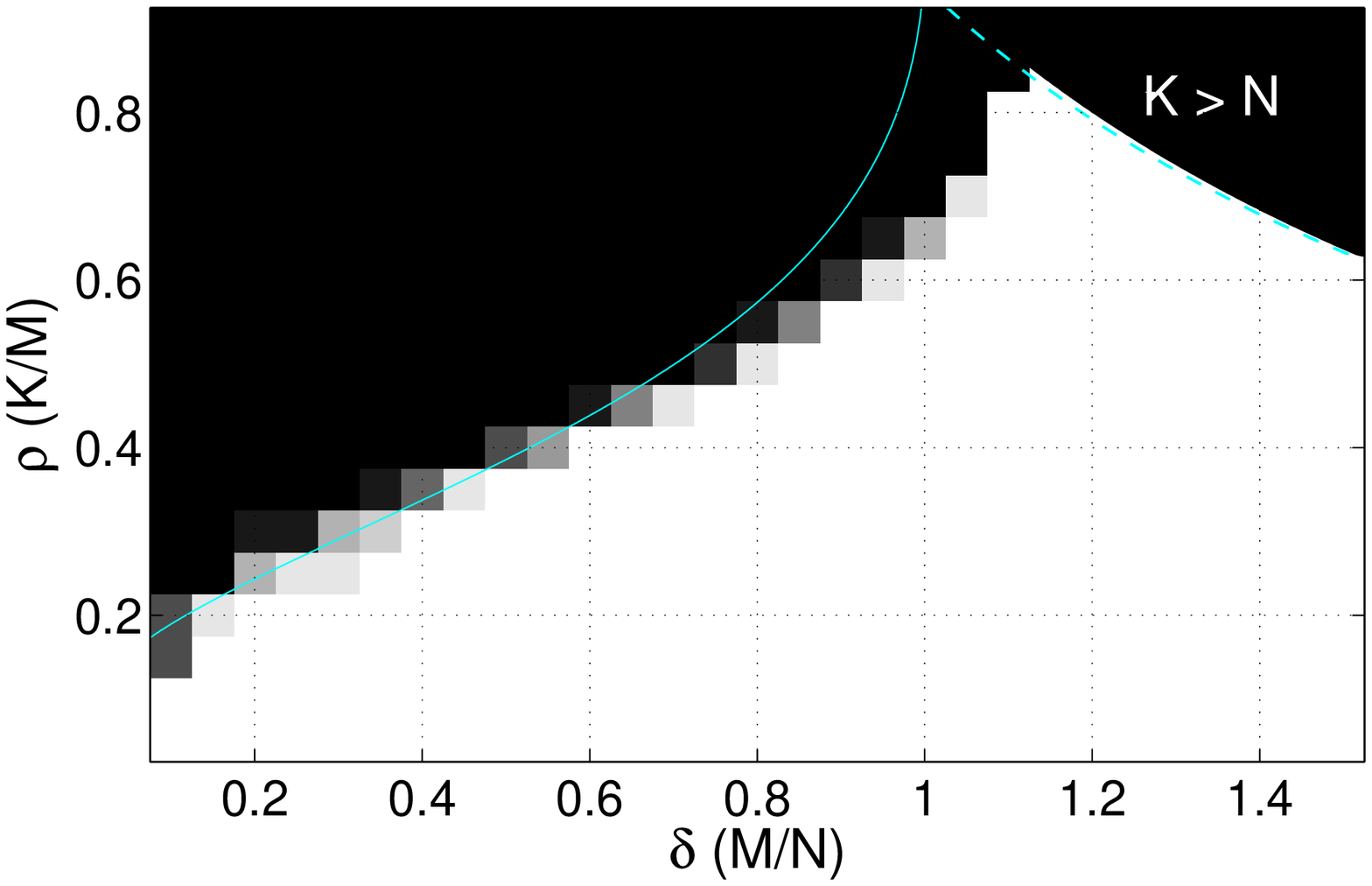}
}
\subfloat{
\includegraphics[width=.23\textwidth]{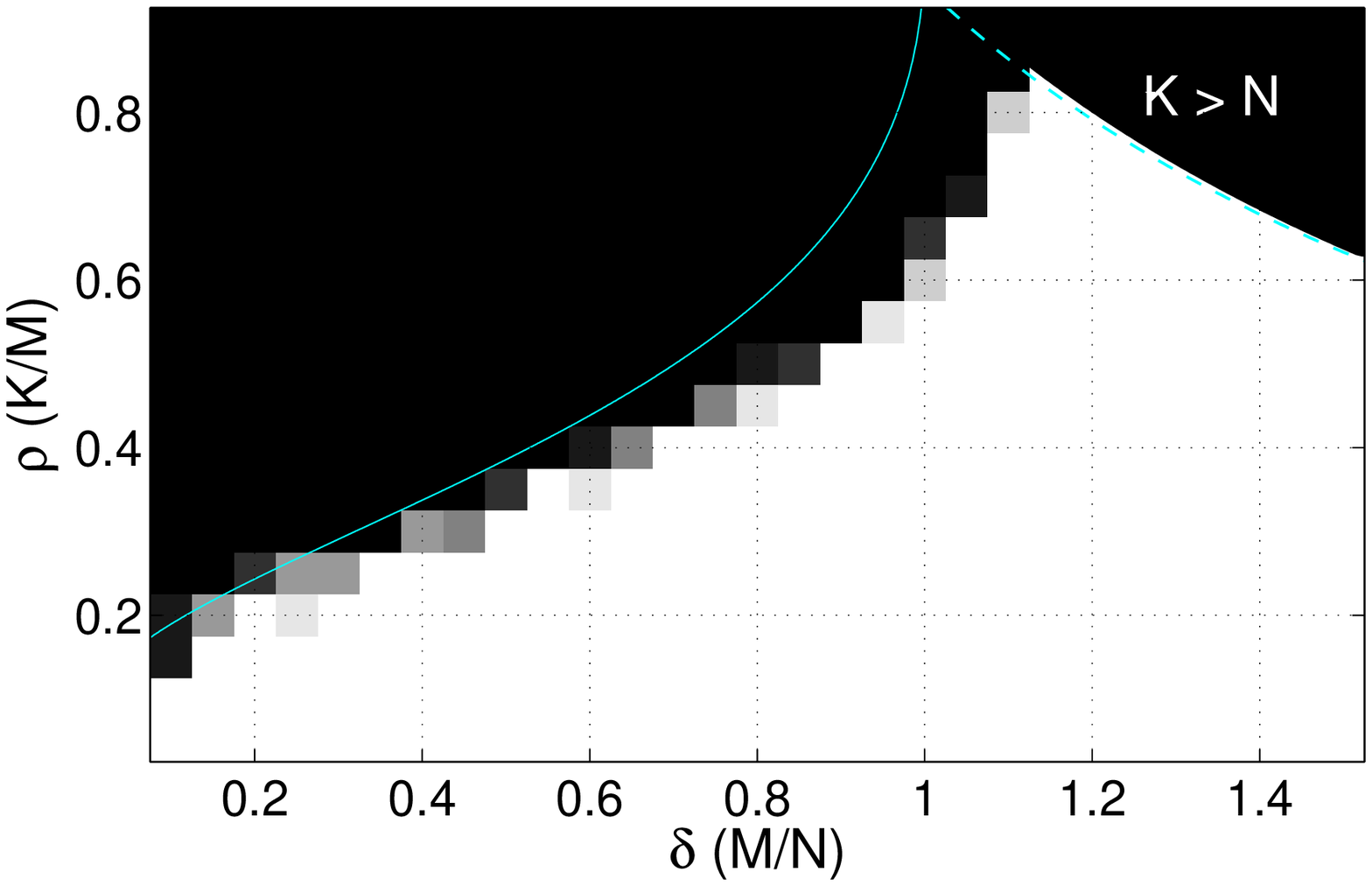}
}
\subfloat{
\includegraphics[width=.23\textwidth]{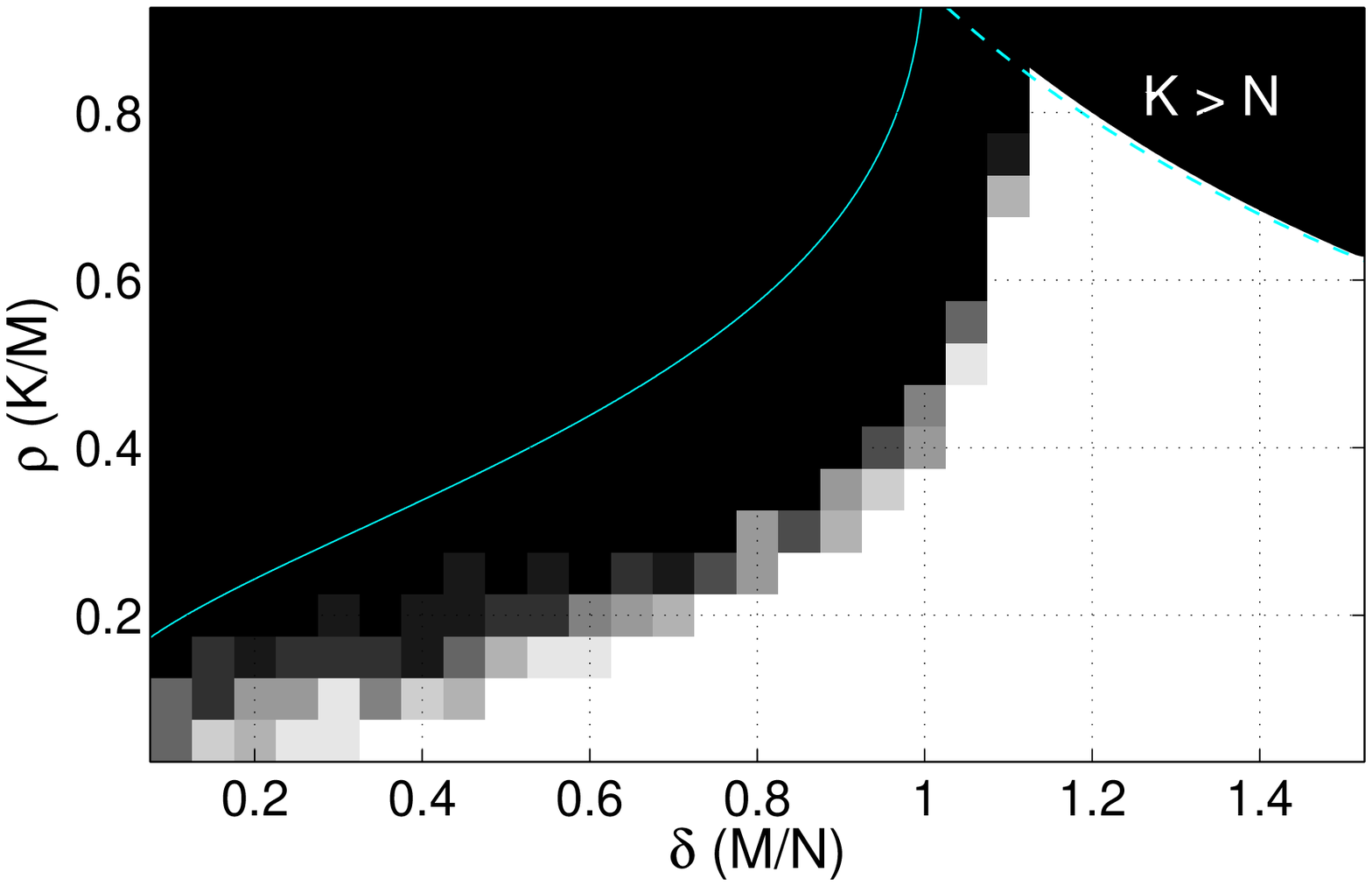}
}
\subfloat{
\includegraphics[width=.23\textwidth]{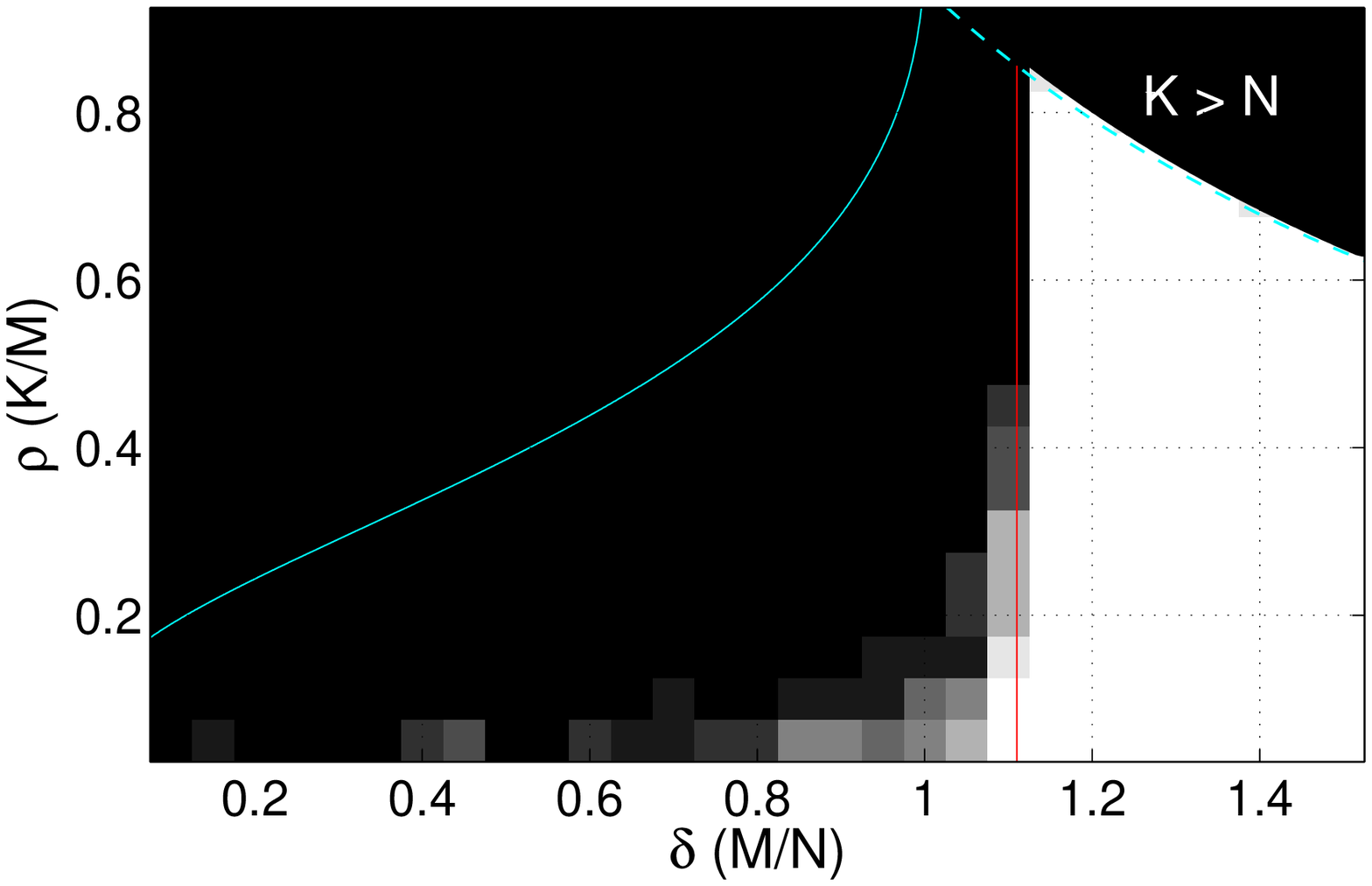}
}
\\
\subfloat{
\rotatebox{90}{{
\begin{minipage}[t]{.14\textwidth}
{\begin{center}
\tiny $\quad\: L=20,\:\sigma=1$
\end{center}}
\end{minipage}
}}
}
\subfloat{
\includegraphics[width=.23\textwidth]{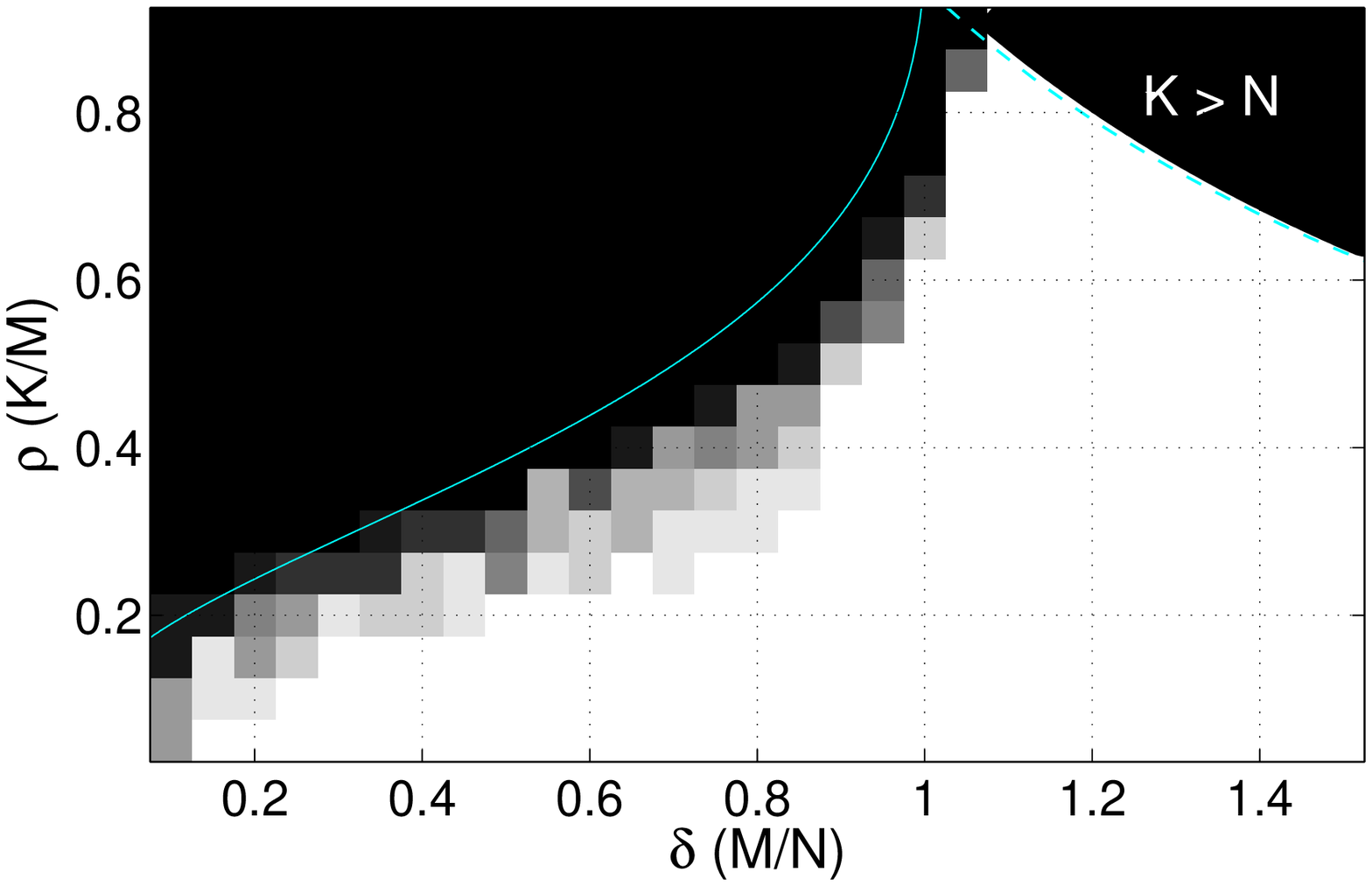}
}
\subfloat{
\includegraphics[width=.23\textwidth]{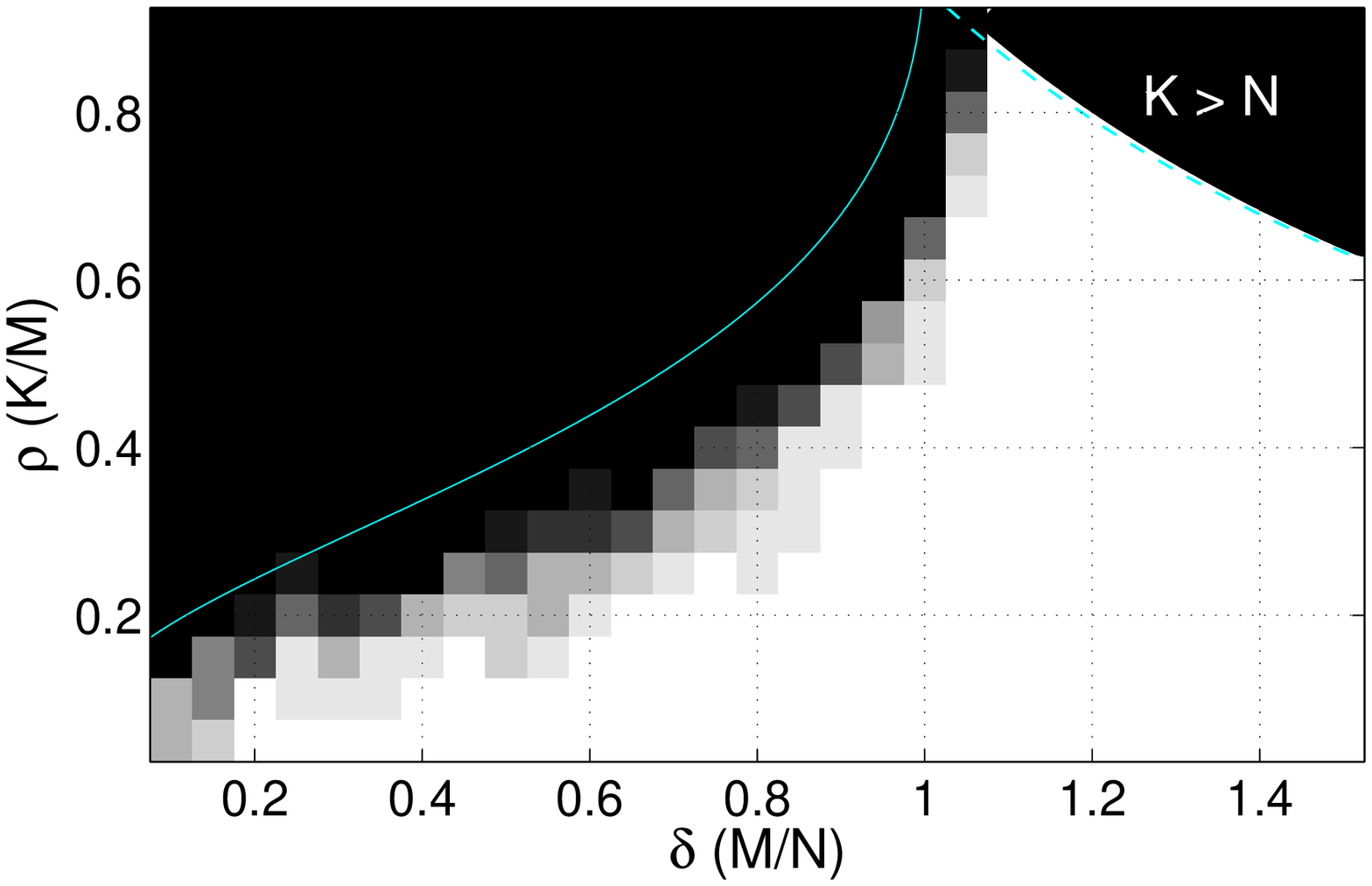}
}
\subfloat{
\includegraphics[width=.23\textwidth]{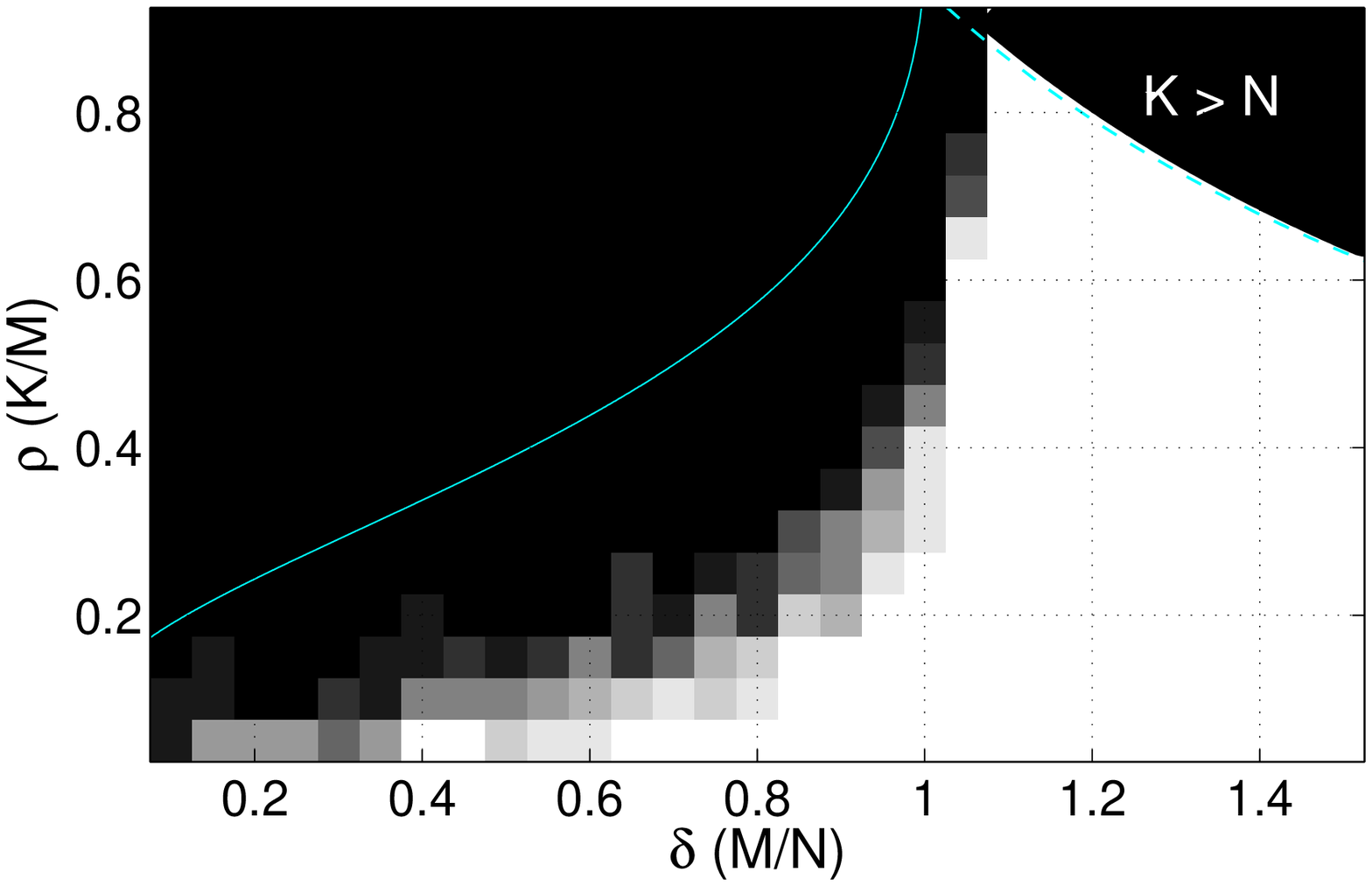}
}
\subfloat{
\includegraphics[width=.23\textwidth]{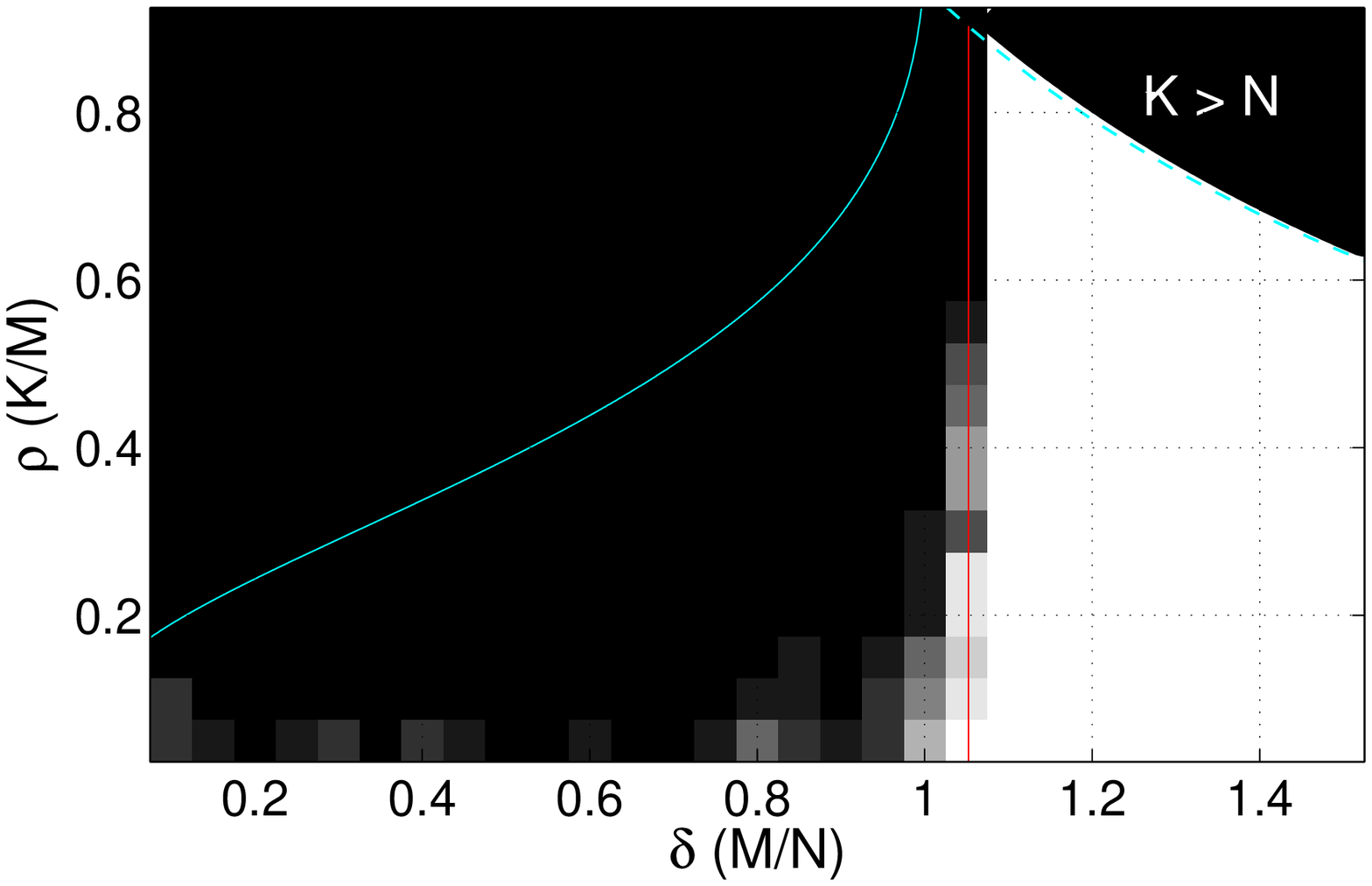}
}
\\
\subfloat{
\rotatebox{90}{{
\begin{minipage}[t]{.14\textwidth}
{\begin{center}
\tiny $\quad\: L=50,\:\sigma=0.1$
\end{center}}
\end{minipage}
}}
}
\subfloat{
\includegraphics[width=.23\textwidth]{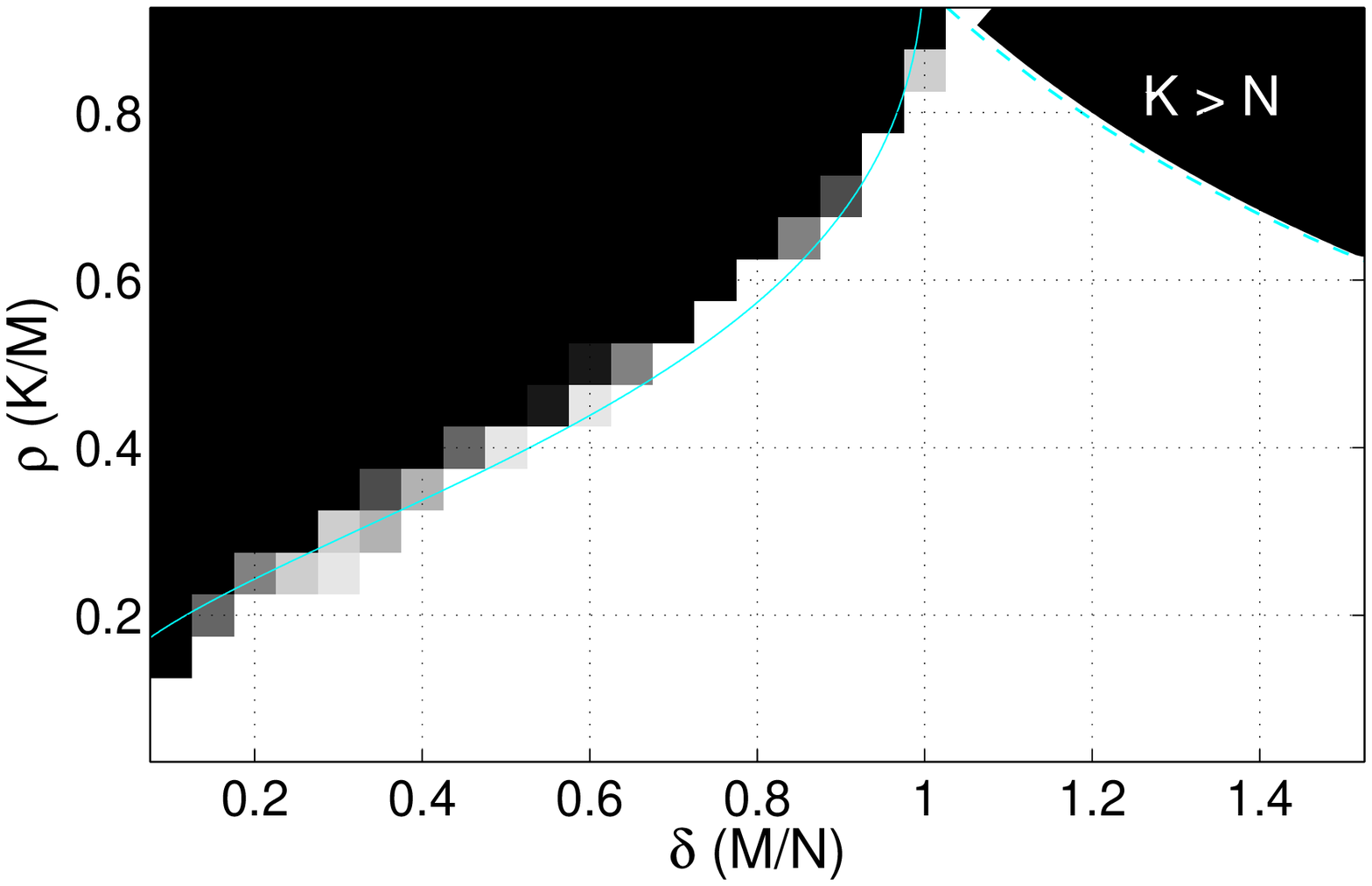}
}
\subfloat{
\includegraphics[width=.23\textwidth]{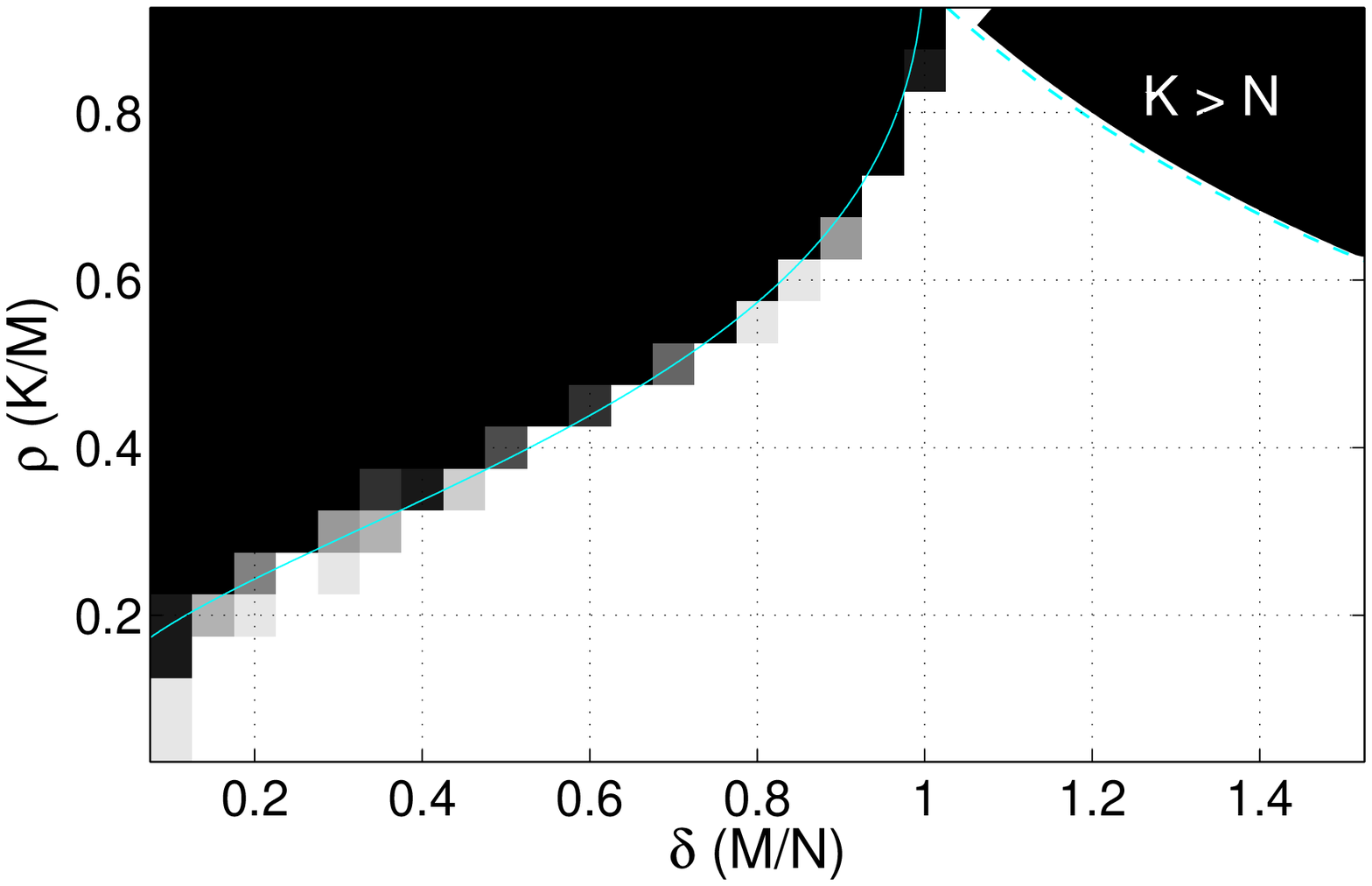}
}
\subfloat{
\includegraphics[width=.23\textwidth]{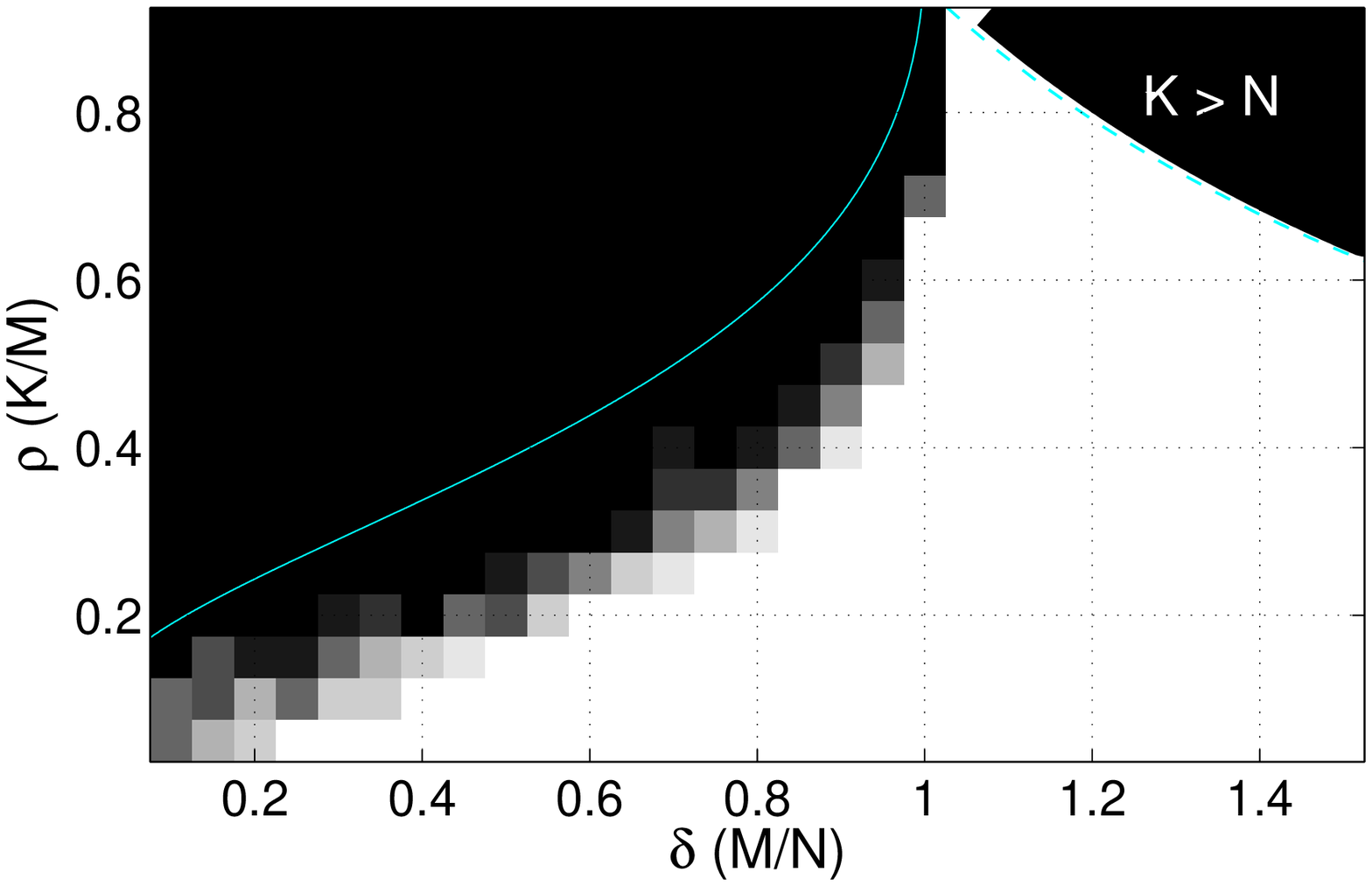}
}
\subfloat{
\includegraphics[width=.23\textwidth]{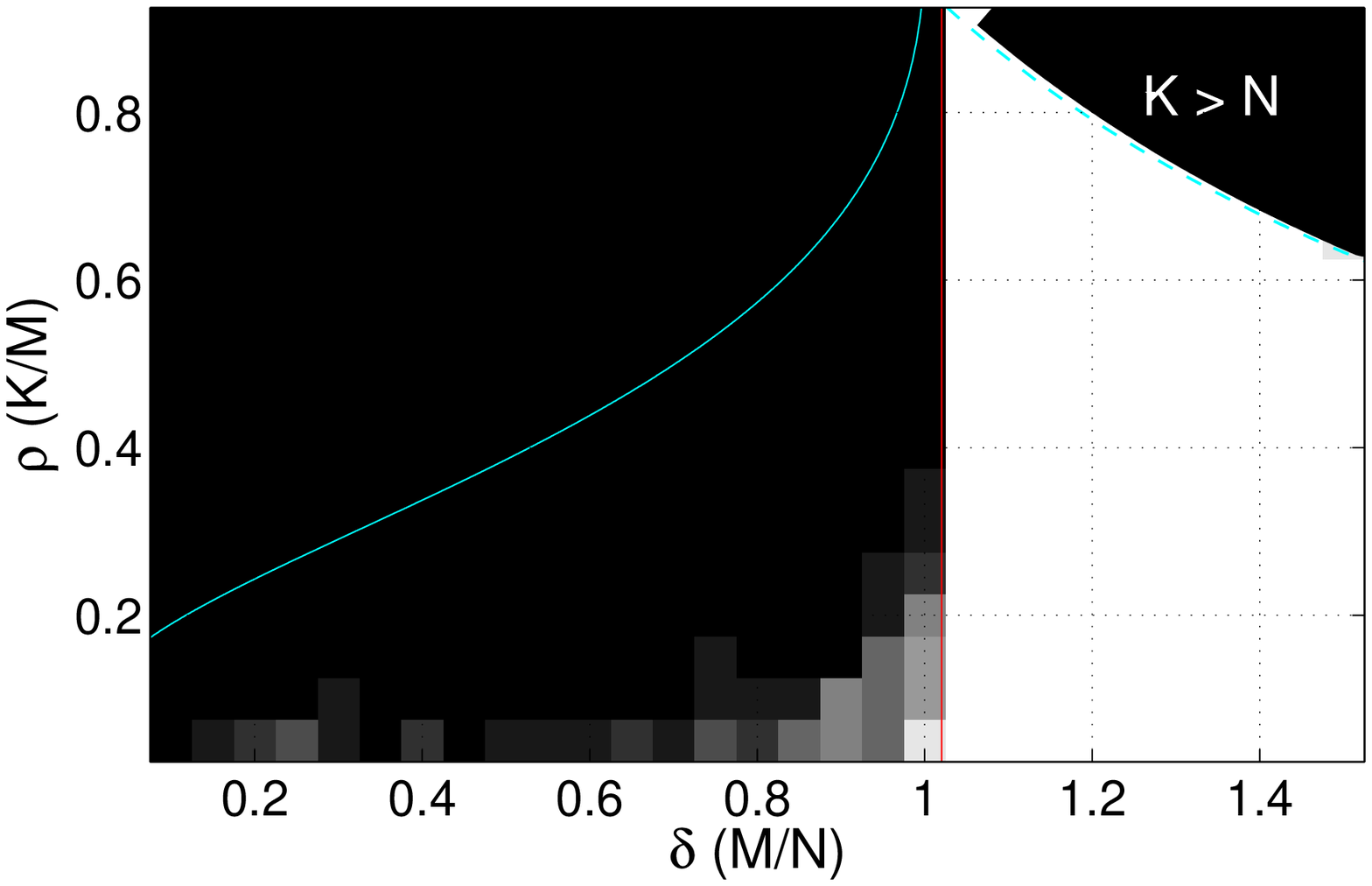}
}
\caption{(\AmpCal) The empirical probability of perfect recovery (colors black to white indicate probabilities 0 to 1) in the complex-valued system for $N=100$ with respect to $\delta \triangleq M/N$ and $\rho \triangleq K/M$. The solid cyan line indicates the Donoho-Tanner phase transition curve for fully calibrated compressed sensing recovery \cite{Donoho2009}. The dashed cyan line indicates the boundary to the unfeasible region where $K>N$. Each row displays the change in recovery performance with increasing phase ambiguity from left to right for a fixed set of $L$ and $\sigma$. The last row shows the performance limit for very large $L$. The red line indicates the bound $\delta=\dcf$ \eqref{eq:deltaBound}. The proposed optimization methods are not required for $\delta\geq\dcf$ since unique recovery is possible through solving the sufficient set of constraint equations.}
\label{fig:complex}
\end{figure*}

The optimization approach {\AmpCal} has been investigated for a real-valued system with strictly positive gains ($\theta_i = 0$) and various experimental results are presented in \cite{Gribonval2012}. However it is also possible to use the same approach for complex-valued systems with different levels of phase variability. Therefore we shall consider the same optimization approach for the calibration of systems with varying characteristics in the distribution of both the phase shifts and the amplitudes of the gains.

\subsubsection{Data Generation}
In order to test the performance of the amplitude calibration approach {\AmpCal}, phase transition diagrams as in compressed sensing recovery are plotted for a signal size $N=100$, with the measurement vectors, $\m_i$, and all the $K$ non-zero entries in the input signals, $\x_\l$, randomly generated from an i.i.d. normal distribution. The positions of the $K$ non-zero coefficients of the input signals, $\x_\l$, are chosen uniformly at random in $\{1, \ldots , N\}$. The magnitude of the gains were generated as having the distribution $\log d_i \sim \mathcal{N}(0,\sigma^{2})$, where $\sigma$ is the  parameter governing the amplitude variability. The phase of the gains are chosen uniformly at random from the range $[0, \; 2\pi p_c)$ where $p_c\in \{0, 1/3, 2/3, 1\}$. Therefore the parameter $p_c$ adjusts the phase variability which is maximized for $p_c=1$.

The signals (and the gains) are estimated for different levels of amplitude variability ($\sigma = 0.1,\; 0.3, \; 1$) and different  numbers of input signals ($L=5,\; 10,\; 20,\; 50$ respectively). For each parameter set, 10 randomly generated set of signals and gains are recovered by an ADMM \cite{Boyd2011} implementation of the proposed optimization {\AmpCal}. 

\subsubsection{Performance Measure}
The perfect reconstruction criterion is selected as $\frac{1}{L}\sum_\l\mu(\xs_\l,\xh_\l)>0.999$, where the absolute correlation factor $\mu(\cdot,\cdot)$ is defined as
\begin{align}
\label{eq:corrfactor}
\mu(\x_1,\x_2) \triangleq \frac{|\x_1\H\x_2|}{\Vert \x_1\Vert_2\Vert \x_2\Vert_2}
\end{align}
so that the global phase and scale difference between the source and recovered signals is ignored.

\subsubsection{Results}
In Figure~\ref{fig:complex}, the performance of {\AmpCal} under different levels of phase and amplitude variability is shown in terms of the probability of recovery (empirically computed through 10 independent simulations for each set of parameters) with respect to $\delta$ and $\rho$ which are defined as
\begin{align}
\delta \triangleq \dfrac{M}{N}, \qquad\qquad
\rho \triangleq \dfrac{K}{M}
\end{align}
The first thing to notice from the results is that the performance in the case of no phase variability ($p_c=0$) is consistent with the results presented in \cite{Gribonval2012} as expected. The effect of increasing phase variability can be observed in the results as $p_c$ increases. Although the performance is acceptable for $p_c$ as high as $2/3$, there is a significant degradation when dealing with very large phase variability ($p_c=1$) so that signal recovery with this approach is impossible regardless of the sparsity, unless a closed form solution is possible ($\delta>\dcf$). This phenomenon can best be observed in the last row of results in Figure~\ref{fig:complex}, where the number of input signals is very large ($L=50$) with respect to the variance in the gain magnitudes ($\sigma=0.1$), but recovery performance still degrades as $p_c$ approaches to 1. Very similar performance for real-valued systems with different sign variability in the gains was observed in experiments for which the results are not shown here.

\begin{figure}[!t]
\centering
\includegraphics[width=.95\columnwidth]{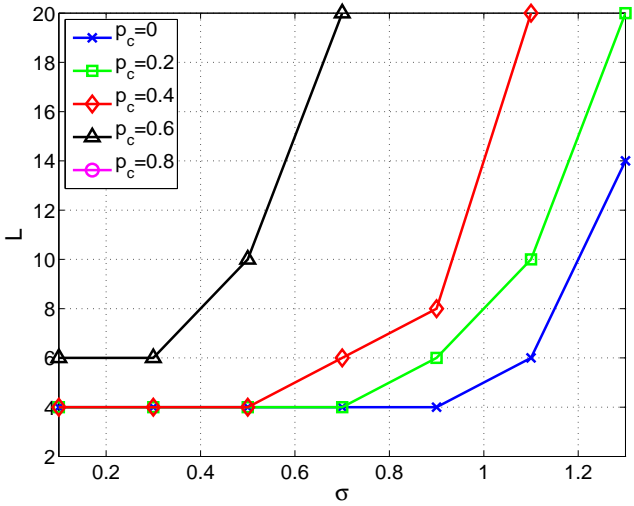}
\caption{(\AmpCal) Minimum number of training signals, $L$, necessary for successfully recovering 97 out of 100 independently generated set of input signals and gains, plotted as a function of amplitude and phase variability ($\sigma$ and $p_c$) when $\rho=0.2$, $\delta=0.8$ and $N=100$. None of the simulated set of number of input signals resulted in a satisfactory rate of successful recovery for the case of $p_c=0.8$.}
\label{fig:psigma}
\end{figure}

The minimum number, $L$, of input signals necessary to obtain a successful recovery in 97 out of 100 independent simulations depends on the amplitude variability, $\sigma$, as well as the phase variability, $p_c$, as can be observed on Figure~\ref{fig:psigma}. For small variability of the phase ($p_c \in \{0, 0.2, 0.4\}$), only a few input signals are sufficient unless there is very large amplitude variability. However as phase variability increases, a very large number of input signals is required regardless of the amplitude variability. The performance can be seen to deteriorate quickly as $p_c$ increases beyond 0.5, to the point that none of the simulated cases results in a satisfactory rate of successful recovery when $p_c=0.8$.

In another experiment which is not shown in the figures, it is observed that the minimum number of input signals required for successful recovery in 97 out of 100 independent simulations remains constant for increasing size of the signal ($L=4$ for all $N=100, 1000, 5000$), when all the other parameters are kept constant ($p_c=0.5$, $\sigma=0.3$, $\rho=0.2$, $\delta=0.8$). 

The diminishing performance of {\AmpCal} as the phase variability, $p_c$, increases can be seen in Figures~\ref{fig:complex} and \ref{fig:psigma}. For the maximum phase variability ($p_c=1$), the results show that {\AmpCal} is not useful since it fails unless $M \geq \dcf N$. Even though this is a drawback of the presented approach, it should be noted that in many practical systems the phase variability is restricted to a small range. Such scenarios represent typical use cases for this optimization approach. We derive in the next section an alternative approach to deal with cases where the phase variability is significant. Further discussion on the performance, benefits and drawbacks of {\AmpCal} can be found in Section~\ref{sec:Conc}.

\subsubsection{Comparison with Related Work}

The only other approach for the gain calibration problem at the time of writing of this paper is the generalized approximate message passing (GAMP) algorithm presented in \cite{Schulke2013}. A performance comparison of the GAMP based algorithm with {\AmpCal} is already presented in \cite{Schulke2013} for real-valued systems with strictly positive gains. In this comparison, the GAMP based method is shown to outperform {\AmpCal} in the setting of random Gaussian measurement matrices, which is not surprising considering the known optimum performance of GAMP in compressed sensing recovery with Gaussian and Bernoulli distributed random matrices. However the GAMP based algorithm is derived by a number of assumptions on the distributions of the gains and the input signals which are not needed for $\l_1$-norm minimization and the robustness of GAMP must be tested against the violation of these assumptions which is out of the scope of this paper. It can be said that, the performance trade off between the GAMP and $\l_1$-norm minimization can also be expected for the calibration algorithms derived from these algorithms. Unfortunately, the comparison for complex-valued gains with different levels of phase variability was simply not possible since the provided implementations of the method in \cite{Schulke2013} only handles the positive real-valued gains and real-valued systems.

\section{Phase Calibration}
\label{sec:PhaseCal}


In this section, we consider the scenario of known gain amplitudes, hence $d_i \m_i$ is simply replaced with $\m_i$ for the rest of the discussions. Therefore we deal with only unknown phase shifts in the measurements.

A similar problem to the phase calibration problem is the phase retrieval problem, in which one tries to retrieve the input signals $\xs$ from the magnitude of the measurements $|y_i|^2=\m_i\H\xs\xs\H\m_i$, $i = 1,\ldots, M$, where $\m_1, \ldots, \m_M$ are vectors of the Fourier basis or Gaussian measurements. Convex methods for phase retrieval have been introduced by Cand\`es \emph{et al.} \cite{Candes2011a} which were based on the lifting schemes introduced earlier by Balan \emph{et al.} \cite{Balan2004,Balan2006,Balan2007}. The extension of these methods when $\xs$ is a sparse vector and the measurements are Gaussian measurements have then been introduced by Ohlsson \emph{et al.} \cite{Ohlsson2011, Ohlsson2012,Ohlsson2013}, who have suggested to use the optimization
\begin{align}
\label{eq:cprl}
\Xh_{\text{CPRL}} = \argmin_{\Z}\quad &\Tr(\Z) + \lambda \Vert \Z \Vert_1 \\
\nonumber 	\text{subject to}\quad & \Z \succcurlyeq 0\\
\nonumber	& |y_i|^2 = \m_i\H\Z\m_i, \quad  i = 1, \ldots, M	,
\end{align}
called the Compressive Phase Retrieval via Lifting (CPRL) where $\Tr(.)$ is the trace of a matrix and the $\ell_1$-norm of the matrix $\Z$ is defined in terms of its scalar entries $\{Z_{i,j} \}$ as
\begin{align}
\label{eq:MatL1norm}
\Vert \Z \Vert_1 \triangleq  \sum_i \sum_j |Z_{i,j}|
\end{align} Provided that sufficiently many measurements are available, the result $\Xh_{\text{CPRL}}$ is shown to coincide with the rank-one, positive semi-definite and sparse matrix $\xs\xs\H$. 

An important observation is that the phase calibration problem becomes identical to the phase retrieval problem when $L=1$, i.e. when there is only a single input signal available. In this case the phase of each observation is simply arbitrary and contains no information to aid the recovery. Therefore only the magnitude measurements $|y_i|^2$ can be used in recovery just as in phase retrieval. 

However when $L>1$, i.e. when more input signals are available, the phase information among the measurements from the same sensor, $y_{i,\l},\: \l=1,\ldots,L$ are correlated, hence we can extend the same approach using quadratic measurements as in \eqref{eq:cprl}, but \emph{without discarding the phase information} in the measurements that can now be relevant for reconstruction. Let us define the cross measurements, $g_{i,k,\l}$ as
\begin{align}
\label{eq:crossmeas}
g_{i,k,\l} &\triangleq y_{i,k}y_{i,\l}\H&  i&=1,\dots, M \\ \nonumber & & k,\l&=1,\dots, L\\
&= e^{j\theta_i} \m_i\H \xs_k \xs_\l\H \m_i e^{-j\theta_i} & & \\
&= \m_i\H \XS_{k,\l} \m_i& \XS_{k,\l}&\triangleq \xs_k \xs_\l\H \in \CoS^{N\times N}
\end{align}
We can also define the joint signal matrix $\XS \in \CoS^{LN\times LN}$
\begin{align}
\XS \triangleq 
\underbrace{
\begin{bmatrix}
\xs_1\\
\vdots\\
\xs_L
\end{bmatrix}
}_{\xs}
 \underbrace{
 \vphantom{\begin{bmatrix}
\xs_1\\
\vdots\\
\xs_L
\end{bmatrix}}
 \begin{bmatrix}
 \xs_1\H & \cdots & \xs_L\H 
 \end{bmatrix}
  }_{\xs\H} =
\begin{bmatrix}
\XS_{1,1} & \cdots & \XS_{1,L}\\
\vdots & \ddots & \vdots\\
\XS_{L,1} & \cdots & \XS_{L,L}
\end{bmatrix}
\end{align}
which is rank-one, hermitian, positive semi-definite and sparse when the input signals, $\xs_\l$, are sparse. Under these conditions, it is possible to apply a similar optimization to \eqref{eq:cprl} in order to retrieve $\XS$ \cite{Bilen2013b}. However the recent results we have obtained through the analysis of this approach suggest that minimization of the trace in \eqref{eq:cprl} provides no benefit to the performance, hence enforcing the sparsity and positive semi-definiteness of $\Z$ is sufficient \cite{Bilen2014}. It is further observed that the convergence speed of the optimization is also improved when the trace is omitted from \eqref{eq:cprl} \cite{Bilen2014}. Therefore we propose to recover the joint matrix $\XS=\xs\xs\H$ with the semi-definite program
\begin{align}
\nonumber {\PhaseCalJ}\textbf{:} \quad\qquad &\\
\label{eq:phasecal}
\Xh = \argmin_{\Z}\quad &\Vert \Z \Vert_1 && \\
\text{subject to}\quad \nonumber & \Z \succcurlyeq 0& &\\
\nonumber & g_{i,k,\l} = \m_i\H\Z_{k,\l}\m_i& i&=1,\dots, M\\
\nonumber & & k,\l&=1,\dots, L
\end{align}
{\PhaseCalJ} enforces sparsity via $\l_1$-norm minimization, while constraining the solution to be a positive semi-definite hermitian matrix that is consistent with the measurements. After the estimation of $\Xh$, the estimated signal $\xh$ (and therefore $\xh_1,\ldots.\xh_L$) is set as $\xh = \R(\Xh,0)$, where the function $\R(.)$ is defined as
\begin{align}
\label{eq:rank1est}
\R(\Z,\phi) \triangleq &\argmin_{\z} \Vert \Z - \z\z\H\Vert_2\\
\nonumber &\text{subject to}\quad \phase(z_{i^*})=\phi
\end{align}
with $\phase(.)$ being the phase of a complex number and $z_{i^*}$ being the first non-zero entry of $\z$. For a hermitian positive semi-definite matrix, the function $\R(.)$ simply outputs the eigenvector corresponding to the largest eigenvalue of the matrix, scaled by the square root of the eigenvalue, phase shifted such that the first non-zero entry of the eigenvector ($z_{i^*}$) has the complex phase $\phi$. Since $\xh$ is defined up to a global phase shift ($\Xh\cong\xh{\xh}\H$), the parameter $\phi$ is simply set as 0. The phases $\theta_i$ can be recovered given $y_{i,\l}$ and $\xh$. 
Even though the input signals are assumed to be sparse, the problem can be modified easily to handle the cases where the signal is sparse in a known domain, $\mathbf{\Psi}$, such that $\Vert \Z \Vert_1$ is replaced with $\Vert \mathbf{\Psi} \Z \mathbf{\Psi}\H\Vert_1$ in \eqref{eq:phasecal}.

\subsection{Experimental Results for Joint Phase Calibration}

\begin{figure*}[!t]
\centering
\subfloat{
\rotatebox{90}{
\begin{minipage}[t]{.01\textwidth}
{ }
\end{minipage}
}}
\subfloat{
\makebox[.23\textwidth][c]{{\footnotesize \qquad$L=1$}}
}
\subfloat{
\makebox[.23\textwidth][c]{{\footnotesize \qquad$L=3$}}
}
\subfloat{
\makebox[.23\textwidth][c]{{\footnotesize \qquad$L=6$}}
}
\subfloat{
\makebox[.23\textwidth][c]{{\footnotesize \qquad$L=10$}}
}
\\
\subfloat{
\rotatebox{90}{{
\begin{minipage}[t]{.14\textwidth}
{\begin{center}
\small {\PhaseCalJ}
\end{center}}
\end{minipage}
}}
}
\subfloat{
\includegraphics[width=.23\textwidth]{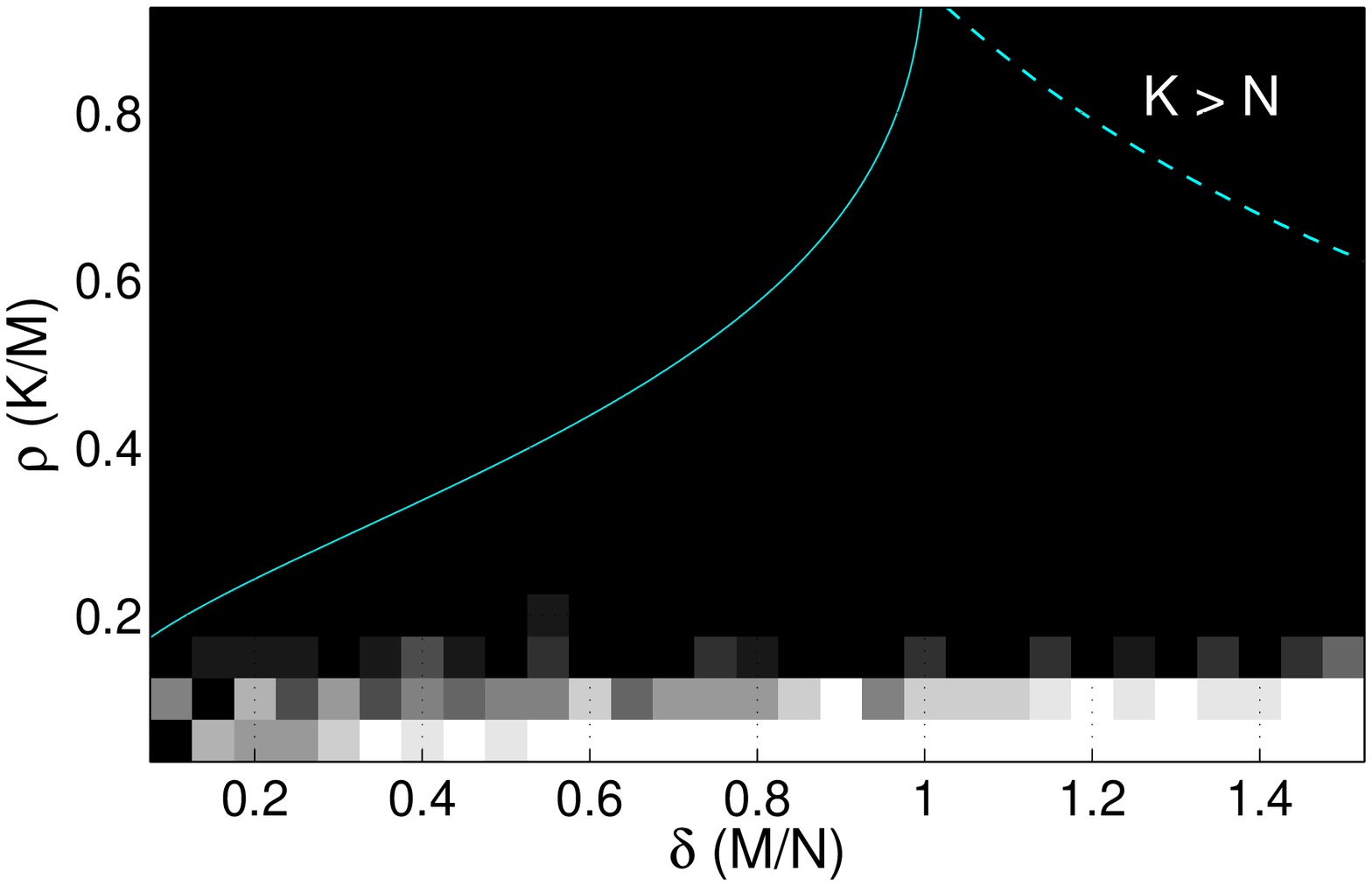}
}
\subfloat{
\includegraphics[width=.23\textwidth]{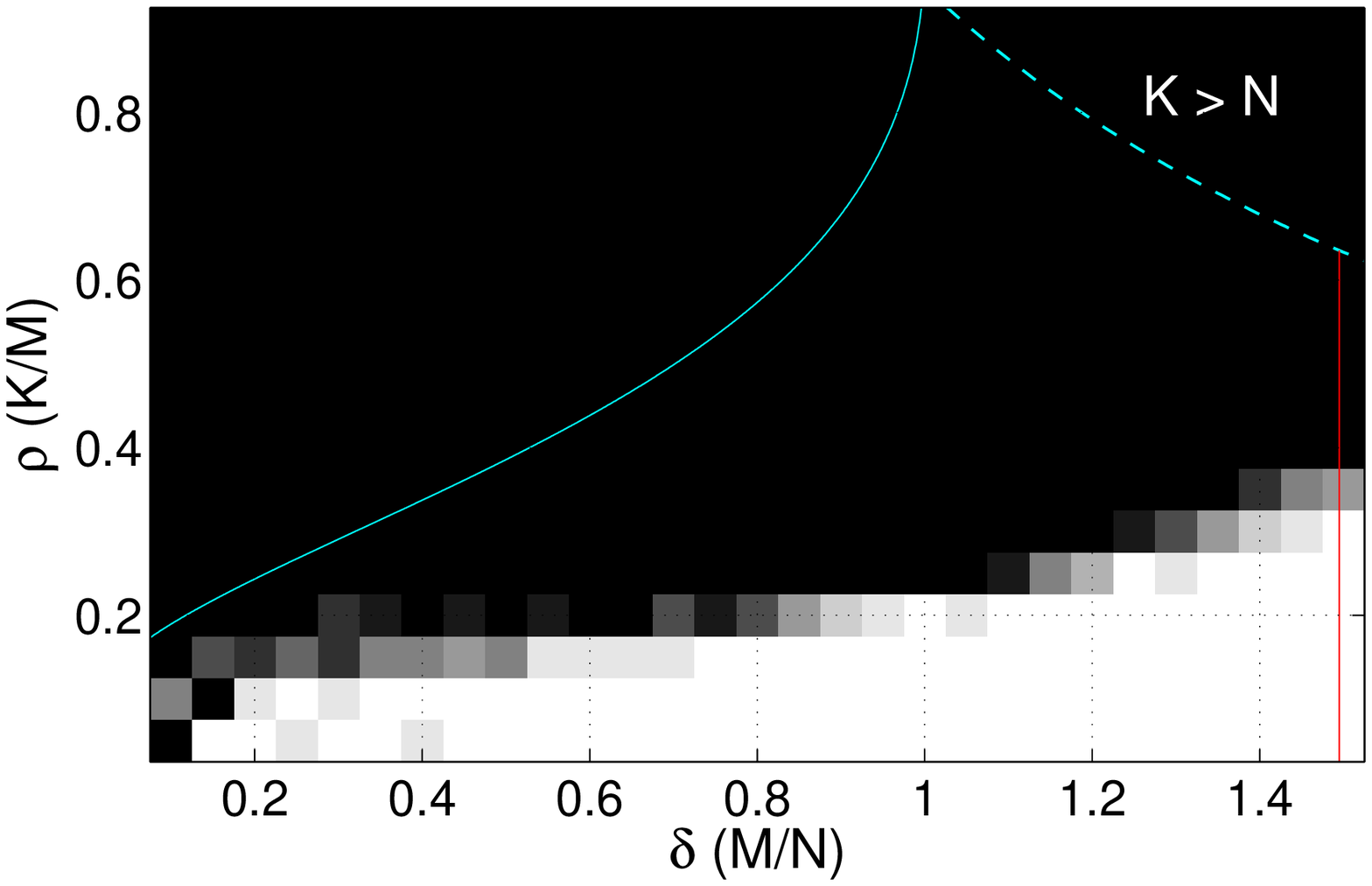}
}
\subfloat{
\includegraphics[width=.23\textwidth]{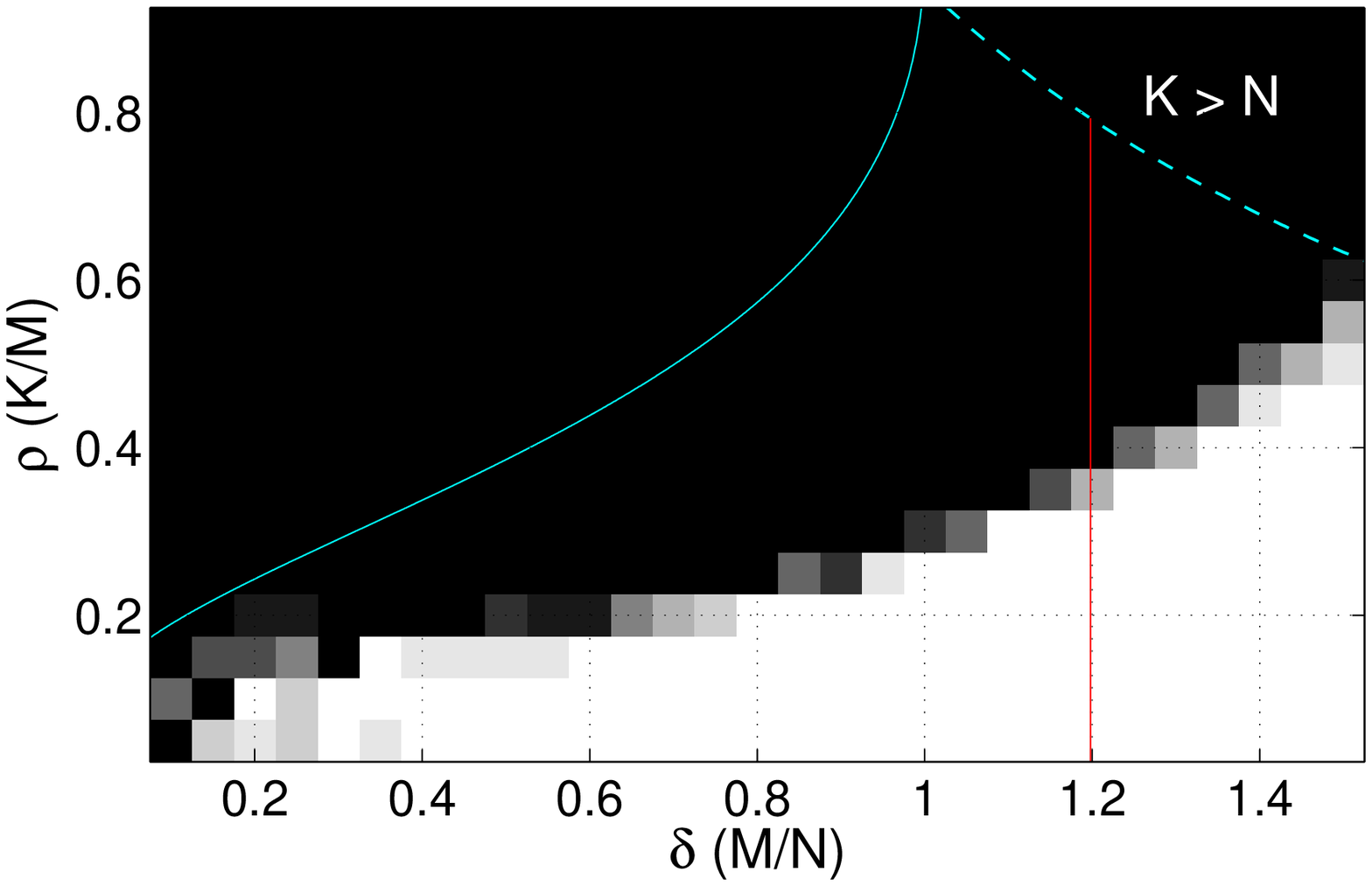}
}
\subfloat{
\includegraphics[width=.23\textwidth]{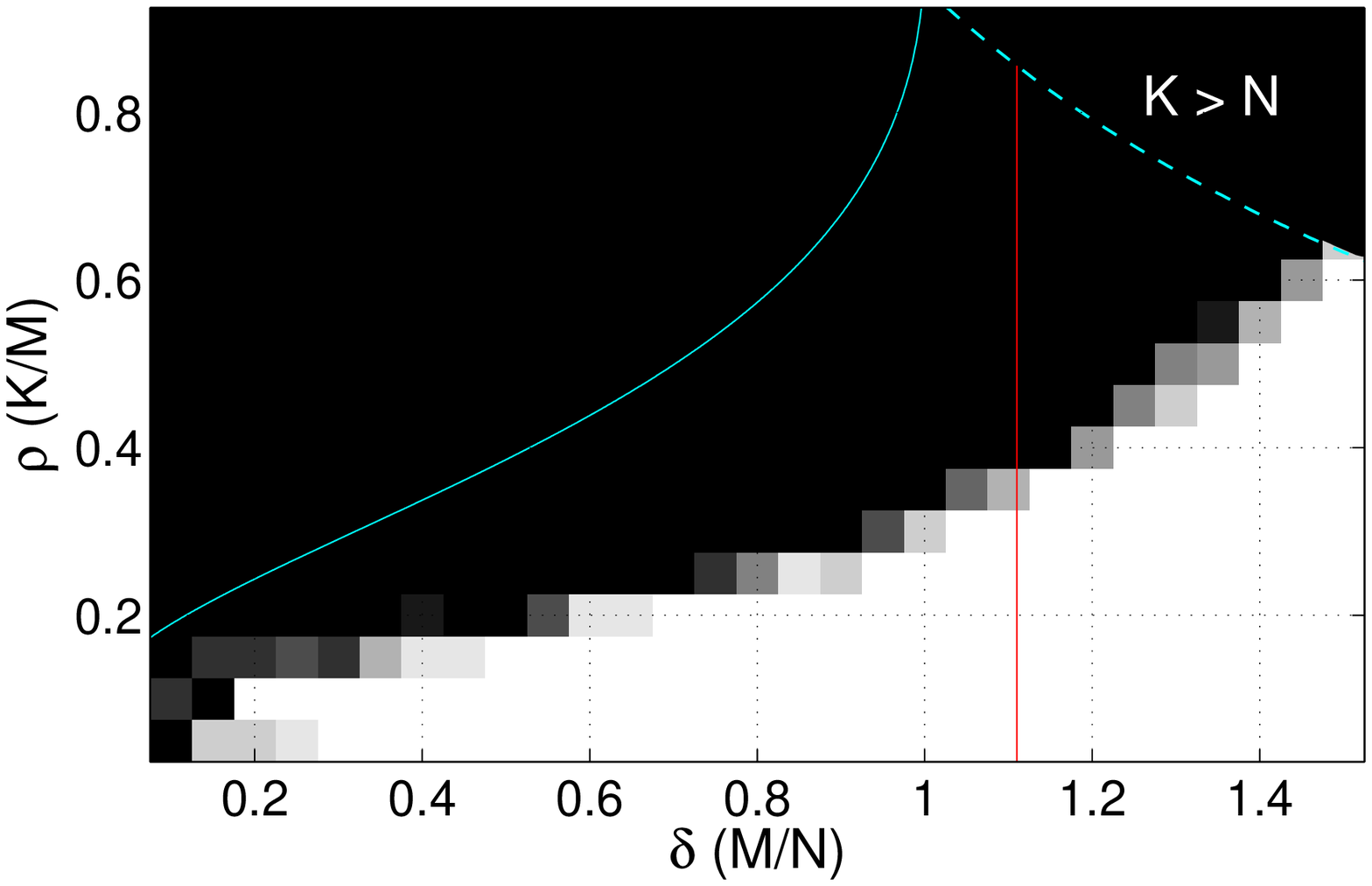}
}
\\
\subfloat{
\rotatebox{90}{{
\begin{minipage}[t]{.14\textwidth}
{\begin{center}
\small {\PhaseCalS}
\end{center}}
\end{minipage}
}}
}
\subfloat{
\includegraphics[width=.23\textwidth]{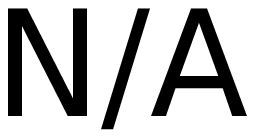}
}
\subfloat{
\includegraphics[width=.23\textwidth]{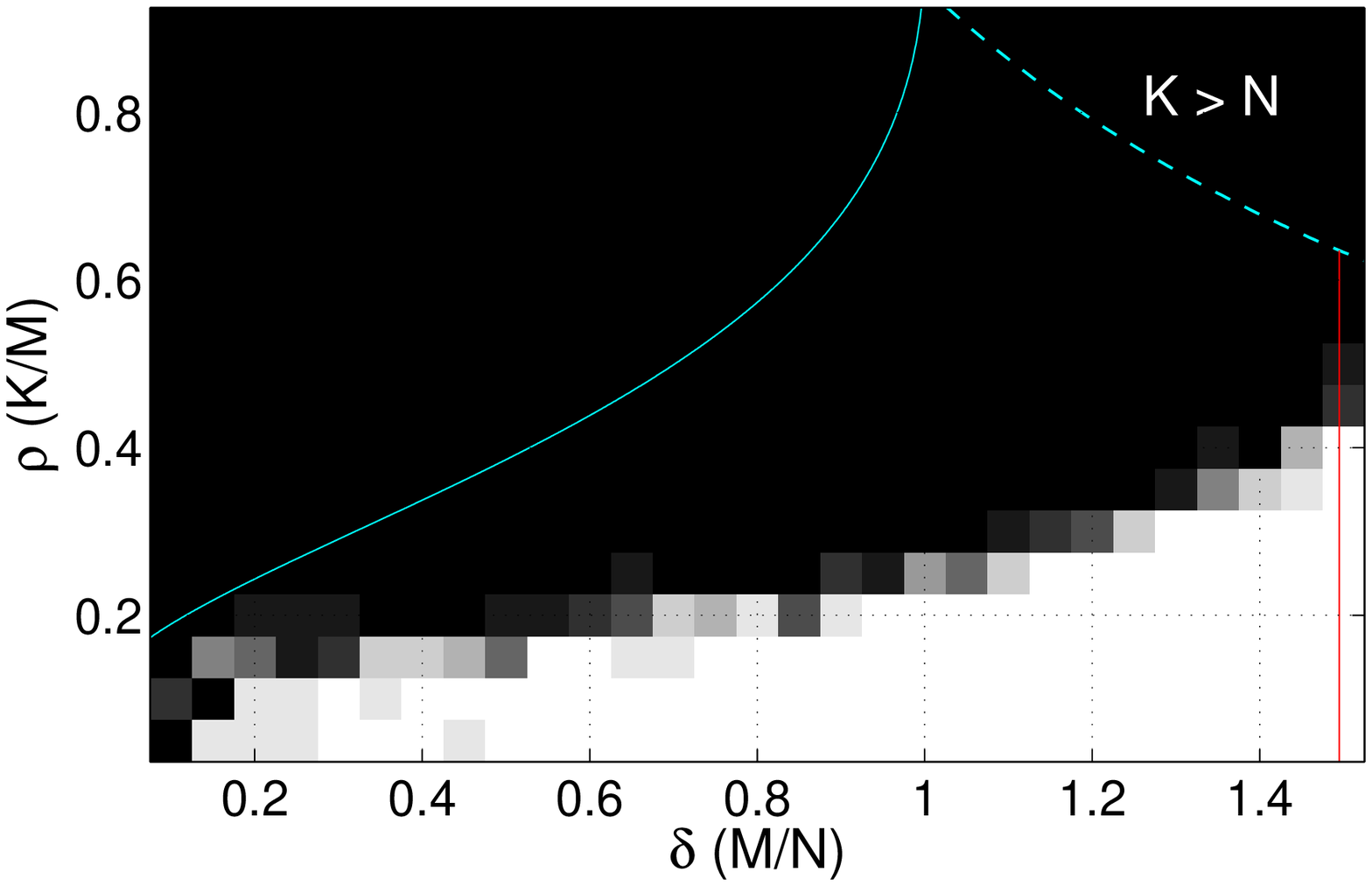}
}
\subfloat{
\includegraphics[width=.23\textwidth]{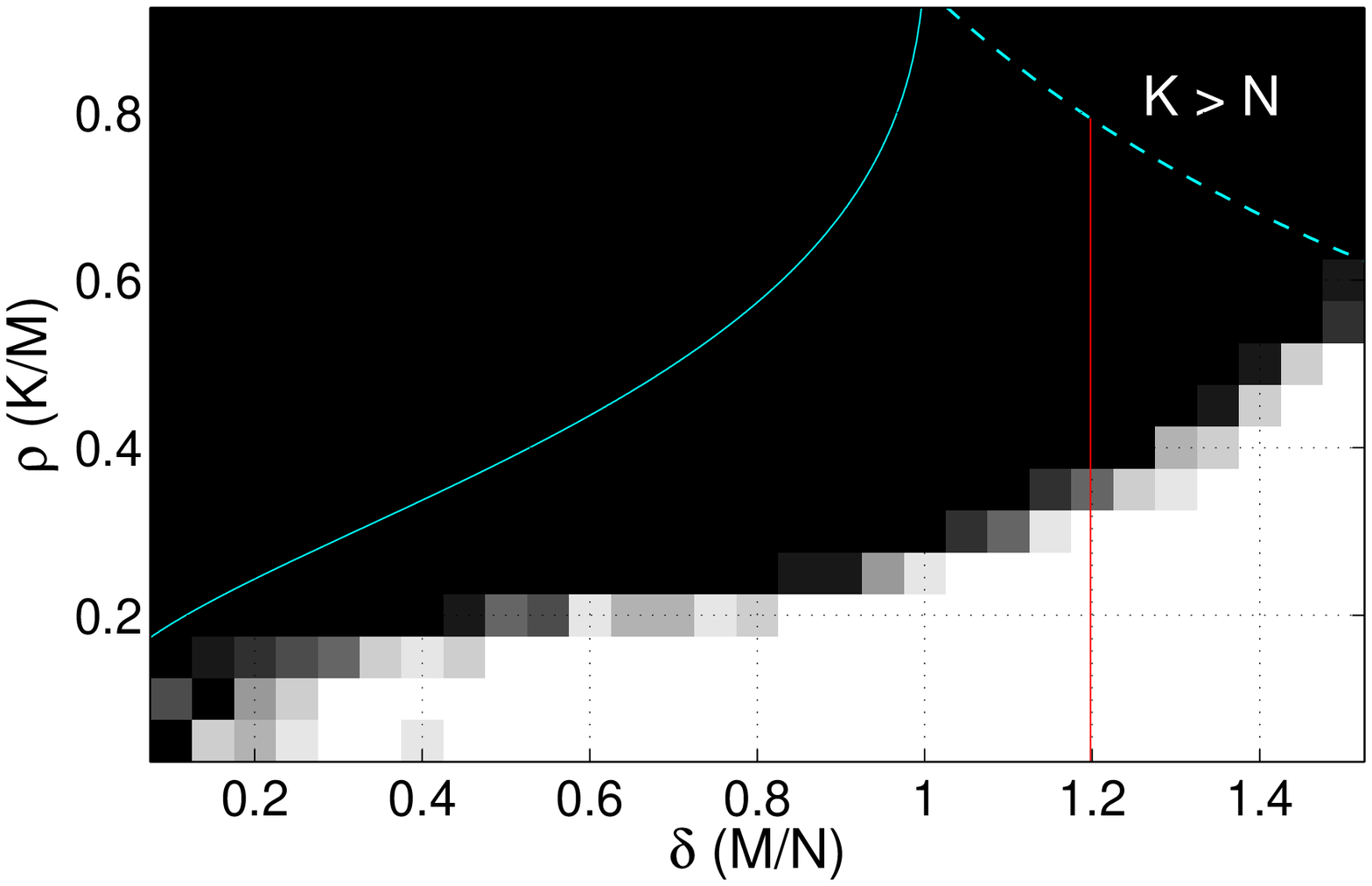}
}
\subfloat{
\includegraphics[width=.23\textwidth]{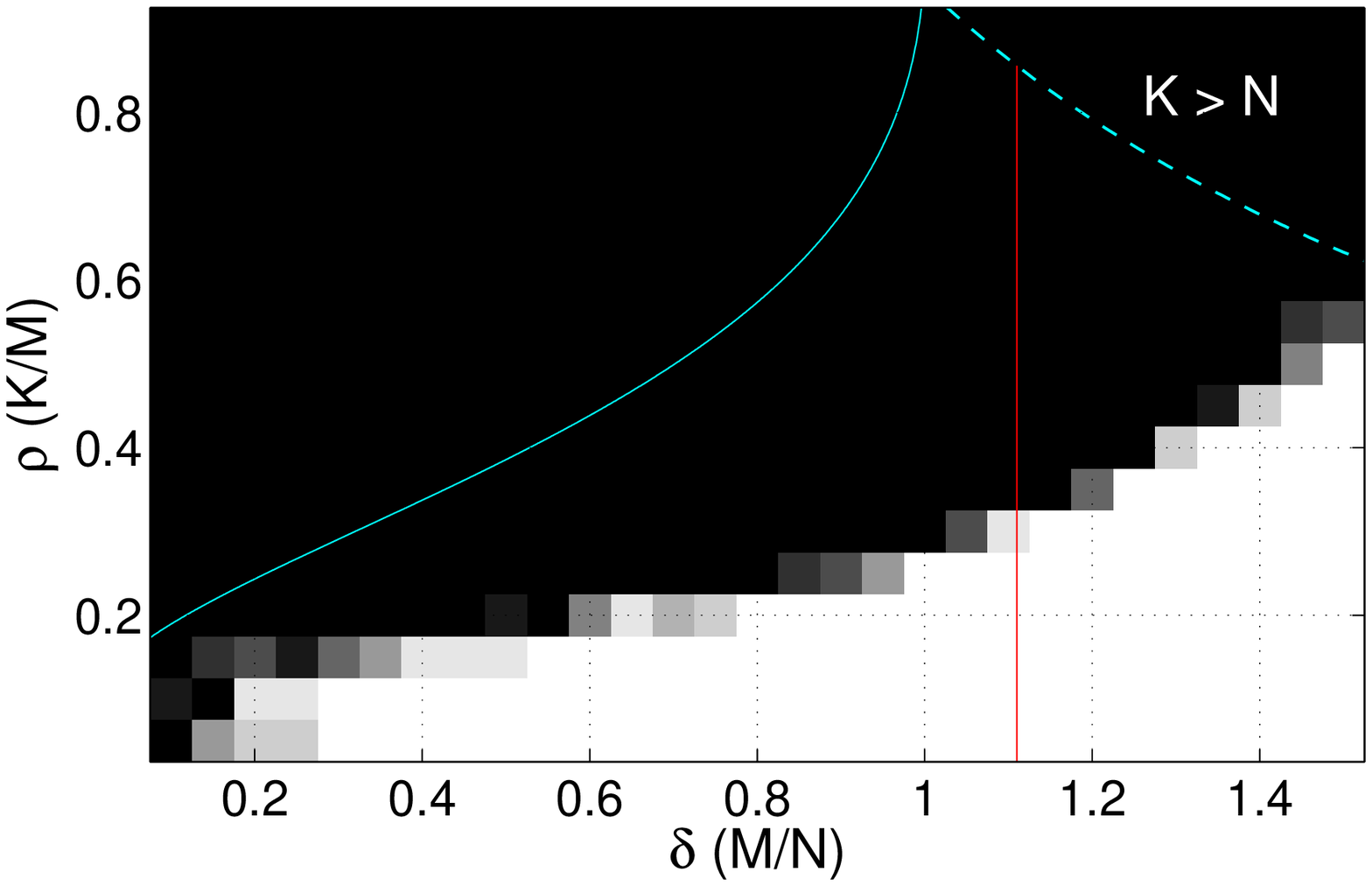}
}
\caption{({\PhaseCalJ} and {\PhaseCalS}) The empirical probability of perfect recovery (colors black to white indicate probabilities 0 to 1) for $N=100$ with respect to $\delta \triangleq M/N$ and $\rho \triangleq K/M$. The solid cyan line indicates the Donoho-Tanner phase transition curve for fully calibrated compressed sensing recovery \cite{Donoho2009}. The dashed cyan line indicates the boundary to the unfeasible region where $K>N$. The rightmost column shows the results with {\PhaseCalS}. The red line indicates the bound $\delta=\dcf$ \eqref{eq:deltaBound}. The proposed optimization methods are not required for $\delta\geq\dcf$ since unique recovery is possible through solving the sufficient set of constraint equations. The optimization {\PhaseCalS} is not considered for $L=1$ since it becomes equivalent to {\PhaseCalJ} for this case.}
\label{fig:complexPT}
\end{figure*}

In order to test the performance of the proposed approach, artificial data are generated as in Section~\ref{sec:AmpCalExp} with no amplitude variability ($d_i=1,\:i=1,\ldots,M$) and maximum phase variability ($\theta_i \in [0,2\pi),\: \text{i.e. } p_c=1$). The signals (and the phase shift parameters) are estimated for the number of input signals $L=1,3,6,10$ with the proposed optimization {\PhaseCalJ} using an ADMM \cite{Boyd2011} algorithm. The performance measure and the perfect reconstruction criteria are also selected as in Section~\ref{sec:AmpCalExp}. 

The probability of recovery of {\PhaseCalJ} with respect to $\delta \triangleq M/N$ and $\rho \triangleq K/M$ is shown in the top row of Figure~\ref{fig:complexPT} for each value of the number of input signals, $L$. It can be observed that the proposed joint recovery methods provide a large improvement in performance to individual optimization (the case with $L=1$), even when there are only few input signals. The performance keeps improving with increasing $L$, although the improvement gets less noticeable as $L$ gets larger up to the point where the performance for $L=10$ is identical to $L=6$. This saturation of performance suggests a lower fundamental limit than the Donoho-Tanner phase transition curve \cite{Donoho2009} that is shown with solid cyan line in Figure~\ref{fig:complexPT}. Such a behavior seems consistent with the theoretical performance of the quadratic basis pursuit for compressive phase retrieval (which is equivalent to phase calibration for $L=1$), for which it has been shown that $K \leq O\left( \sqrt{\frac{M}{\log N}} \right)$ is sufficient for perfect recovery with high probability rather than  $O\left( \frac{M}{\log N} \right)$ as in compressed sensing recovery \cite{Li2012}.

\begin{figure}[!t]
\centering
\includegraphics[width=.95\columnwidth]{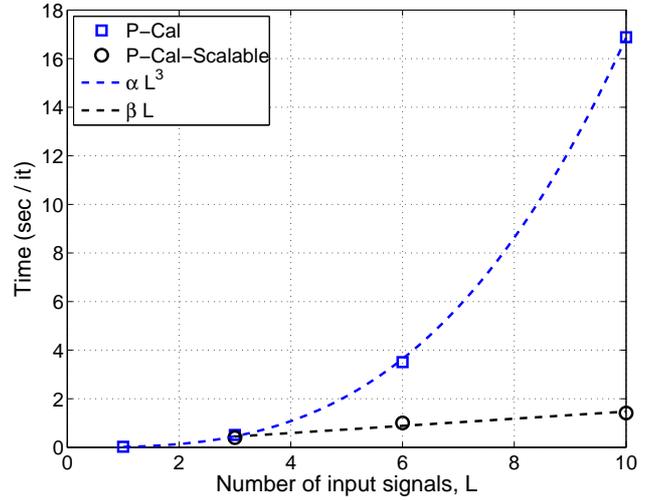}
\caption{({\PhaseCalJ} and {\PhaseCalS}) The average time it takes per iteration for the ADMM implementations of {\PhaseCalJ} and {\PhaseCalS} during recovery for  $N=100$, $\delta=1$, $\rho=0.25$. The time per iteration is computed by averaging both among the iterations and among 10 different realizations of the same experiment.}
\label{fig:pcal_time}
\end{figure}

The complexity of {\PhaseCalJ} is expected to be in the order of $O(L^3 N^3)$ since it is a semi-definite program dealing with a positive semi-definite matrix of size $LN\times LN$. This is empirically demonstrated in Figure~\ref{fig:pcal_time}, in which average time per iteration of the {\PhaseCalJ} algorithm is shown to match with a curve of the form $\alpha L^3$ where $\alpha$ is a constant. The size of the variables also imply that the memory requirements are of the order $O(L^2 N^2)$. This significant grow of complexity and memory limits in practice the number of input signals that can be processed jointly in the optimization. Hence, a modified version of {\PhaseCalJ} is presented in Section~\ref{sec:PhaseCalSc} in order to provide a more scalable performance with respect to $L$. 

\section{Scalable Phase Calibration}
\label{sec:PhaseCalSc}

\begin{figure}[!t]
\centering
\psscalebox{0.95}{
\psset{unit=0.36cm}
\input{matrix.tex}
}
\caption{The sub-matrices $\Xs_1,\ldots,\Xs_L$ within the matrix $\X$.}
\label{fig:mat_str}
\end{figure}

The drawback of the quadratic formulation in {\PhaseCalJ} is that the problem size grows with the square of the number of unknowns in the original problem ($\Xh \in \CoS^{LN\times LN}$) instead of the linear relation in {\AmpCal} ($\xh \in \CoS^{NL}$). In this section we propose an alternative approach for the phase calibration in order to remedy this issue partially so that the problem size becomes linear in the number of input signals, $L$. 

A simple approach to reduce the optimization complexity of {\PhaseCalJ} is recovering $\XS_{\l,\l}$ individually by enforcing the sparsity, the positive semi-definiteness and the constraints for cross measurements $g_{i,\l,\l}$ for $i=1,\ldots,M$ and $\l=1,\ldots,L$, which would result in an optimization with complexity linear in $L$. However this would also mean not benefiting from the joint optimization and the performance gain of {\PhaseCalJ} for $L>1$ shown in Figure~\ref{fig:complexPT}.

It can be observed that the constraints enforced by the cross measurements in \eqref{eq:crossmeas} form a significantly overcomplete set of equations. In fact, given $2ML$ cross measurements (namely $g_{i,\l,\l}$ and $g_{i,\l,(\l)_L+1}$ for $i=1,\ldots,M$ and $\l=1,\ldots,L$ where $(\l)_L \triangleq \l \pmod L$), it is possible to deduce the rest of the cross measurements (and therefore all of the original measurements, $y_{i,\l}$, up to a global phase shift). The rank-one positive semi-definite matrix $\XS$ can also be fully defined given the entries $\XS_{\l,\l}$ and $\XS_{\l,(\l)_L+1}$ for $\l=1,\ldots,L$. In light of these observations, one can aim to recover a subset of $\XS$ using only a subset of the cross measurements, thus reducing the optimization complexity. For this purpose, we can aim to construct the sub-matrices $\Xs_{\l}$ of $\XS$, defined as
\begin{align}
\label{eq:submat}
\Xs_{\l} &\triangleq 
\begin{bmatrix*}[l]
\XS_{\l,\l} & \XS_{\l,(\l)_L+1}\\
\XS_{(\l)_L+1,\l} & \XS_{(\l)_L+1,(\l)_L+1}
\end{bmatrix*}\quad \l=1,\dots, L
\end{align}
A depiction of the sub-matrices with respect to $\X$ can be seen in Figure~\ref{fig:mat_str}. Similarly to $\XS$, $\Xs_\l$ are also rank-one, hermitian, positive semi-definite and sparse matrices. Therefore, an alternative to the optimization {\PhaseCalJ} is solving
\begin{align}
\nonumber {\PhaseCalS}\textbf{:}\qquad\quad & \\
\label{eq:phasecal_fast}
\nonumber \begin{array}{c}
\X_{1,1},\ldots,\X_{L,L}\\
\X_{1,2},\ldots,\X_{L,1}
\end{array} =\quad & \\
\argmin_{\substack{\Z_{1,1}\dots\Z_{L,L}\\ \Z_{1,2}\dots\Z_{L,1}}} \sum\limits_{k=1}^L \left( \Vert\Z_{k,k}\Vert_1 \right. & + \left. \Vert\Z_{k,(k)_L+1}\Vert_1 \right) \\
\nonumber \text{subject to} \qquad g_{i,\l,\l} = & \m_i\H\Z_{\l,\l}\m_i\qquad\qquad\: i=1,\dots, M  \\
\nonumber  g_{i,\l,(\l)_L+1} = & \m_i\H\Z_{\l,(\l)_L+1}\m_i \qquad\l=1,\dots, L \\
\nonumber \left[ \begin{array}{l}
\Z_{\l,\l}\\
\Z_{\l,(\l)_L+1}\H
\end{array} \right. & \left. \begin{array}{l}
\Z_{\l,(\l)_L+1}\\
\Z_{(\l)_L+1,(\l)_L+1}
\end{array} \right] \succcurlyeq 0
\end{align}
After performing this optimization, the estimated signals are set such that
\begin{align}
\begin{bmatrix*}[l]
\xh_1 \\ \v_2
\end{bmatrix*} = \R(\Xsh_1,0)
\end{align}
and
\begin{align}
\begin{bmatrix*}[l]
\xh_\l \\ \v_{(\l)_L+1}
\end{bmatrix*} = \R(\Xsh_\l,\phase(v_{\l,i^*})),\quad \l=2,\ldots,L
\end{align}
in which $v_{\l,i^*}$ is the first non-zero entry of the vector $\v_\l$.


The optimization {\PhaseCalS} deals with $L$ matrices of size $2N\times 2N$ instead of a single $LN\times LN$ matrix with $2ML$ constraints instead of $ML^2$ with respect to {\PhaseCalJ}. Hence the expected memory requirement is reduced from $O(L^2 N^2)$ to $O(L N^2)$ while complexity is reduced from $O(L^3 N^3)$ to $O(LN^3)$. The joint recovery characteristic of {\PhaseCalJ} is still preserved unlike individual recovery of each input signal. A comparison can be seen in Table~\ref{tab:comparison} and an empirical demonstration of the order of complexity is shown in Figure~\ref{fig:pcal_time}. 

\begin{table}
\caption{Comparison between the proposed methods and the independent recovery of $\XS_{\l,\l}$ for $\l=1,\ldots,L$, for the reconstruction of $L$ input signals of size $N$}
\centering
\begin{tabular}{L{.14\columnwidth} | C{.30\columnwidth}C{.17\columnwidth}C{.17\columnwidth}}
\label{tab:comparison}
 & $\#$ of Constraints \&\newline $\#$ of Unknowns & Memory\newline Requirement & Complexity\\
\hline
Individual Recovery & $ML$ const. \newline $L$ $N\times N$ mat. & $O(LN^2)$ & $O(LN^3)$\\
& & &\\
{\PhaseCalJ} & $ML^2$ const. \newline $NL \times NL$ mat. & $O(L^2 N^2)$ & $O(L^3 N^3)$\\
& & &\\
{\PhaseCalS} & $2ML$ const. \newline $2L$ $N \times N$ mat. & $O(LN^2)$ & $O(LN^3)$
\end{tabular}
\end{table}

\subsection{Experimental Results for Scalable Phase Calibration}

In order to measure the performance of {\PhaseCalS}, an experimentation procedure identical to {\PhaseCalJ} is followed to generate the empirical probability of recovery with respect to $\delta \triangleq M/N$ and $\rho \triangleq K/M$, for the number of input signals $L=6$ and $L=10$. The performance of {\PhaseCalS} is shown in the bottom row of  Figure~\ref{fig:complexPT}. 

The recovery performance of {\PhaseCalS} can be seen to be almost identical to {\PhaseCalJ}, especially for $\delta<\dcf$ for which a closed form solution is not possible and optimization is needed. However the complexity of {\PhaseCalS} is significantly lower than {\PhaseCalJ} as stated earlier, which can also be observed in Figure~\ref{fig:pcal_time}. The memory requirements for {\PhaseCalS} is also reduced as shown in Table~\ref{tab:comparison}.

\section{Complete Calibration}
\label{sec:CompCal}


\begin{figure*}[!t]
\centering
\subfloat{
\makebox[.33\textwidth][c]{{\footnotesize $\qquad \sigma=0.1$}}
}
\subfloat{
\makebox[.33\textwidth][c]{{\footnotesize $\qquad \sigma=0.3$}}
}
\subfloat{
\makebox[.33\textwidth][c]{{\footnotesize $\qquad \sigma=0.5$}}
}
\\
\subfloat{
\includegraphics[width=.33\textwidth]{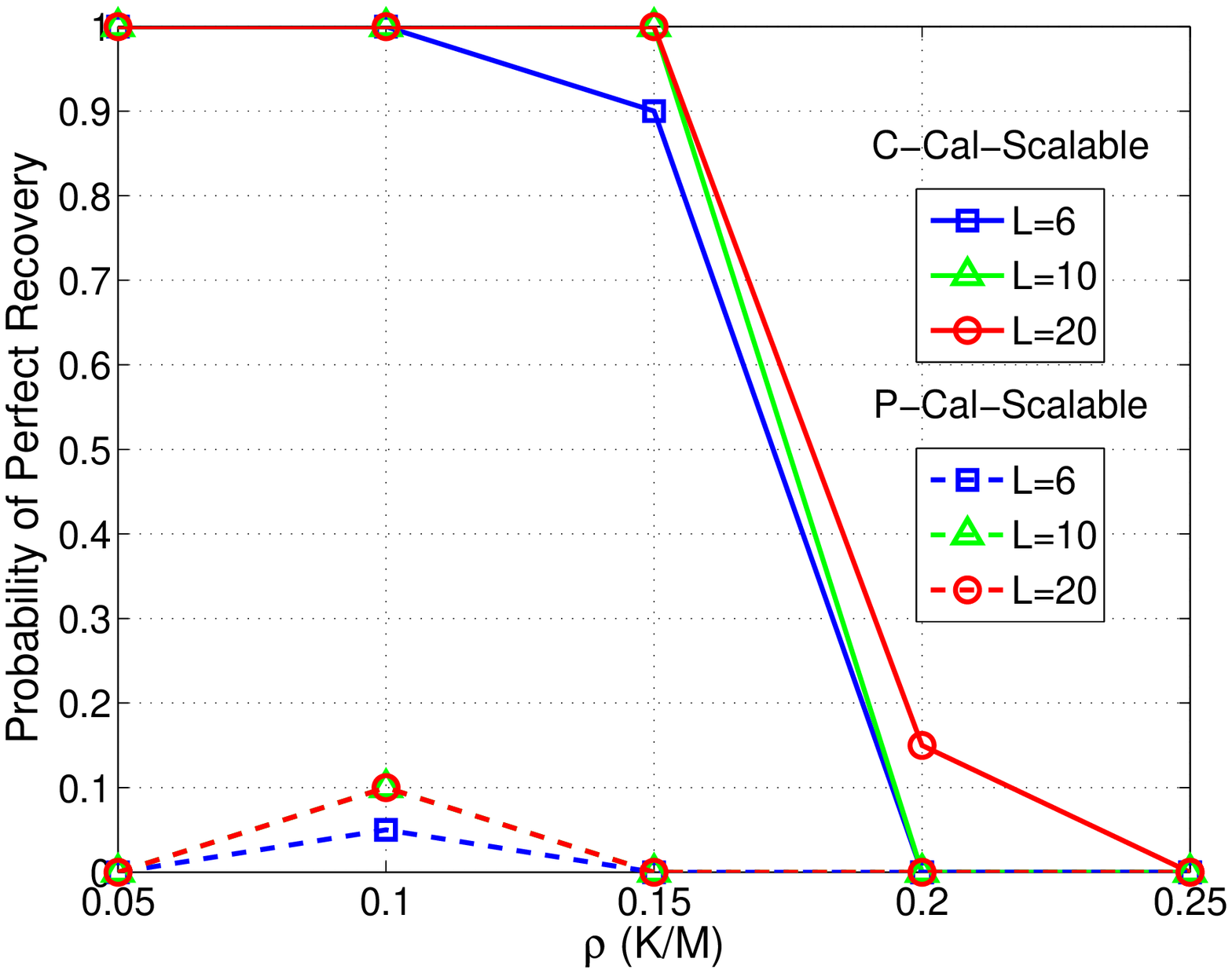}
}
\subfloat{
\includegraphics[width=.33\textwidth]{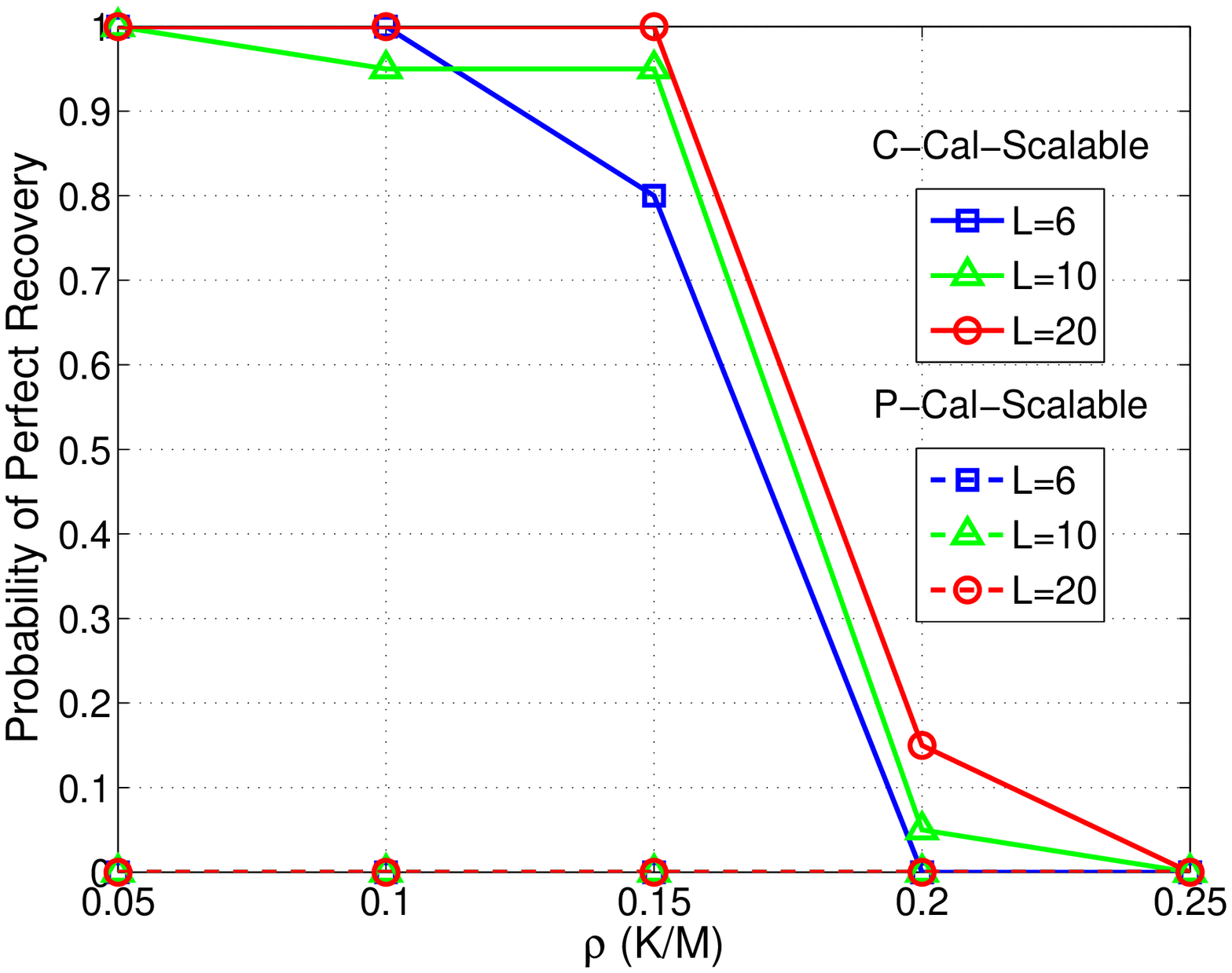}
}
\subfloat{
\includegraphics[width=.33\textwidth]{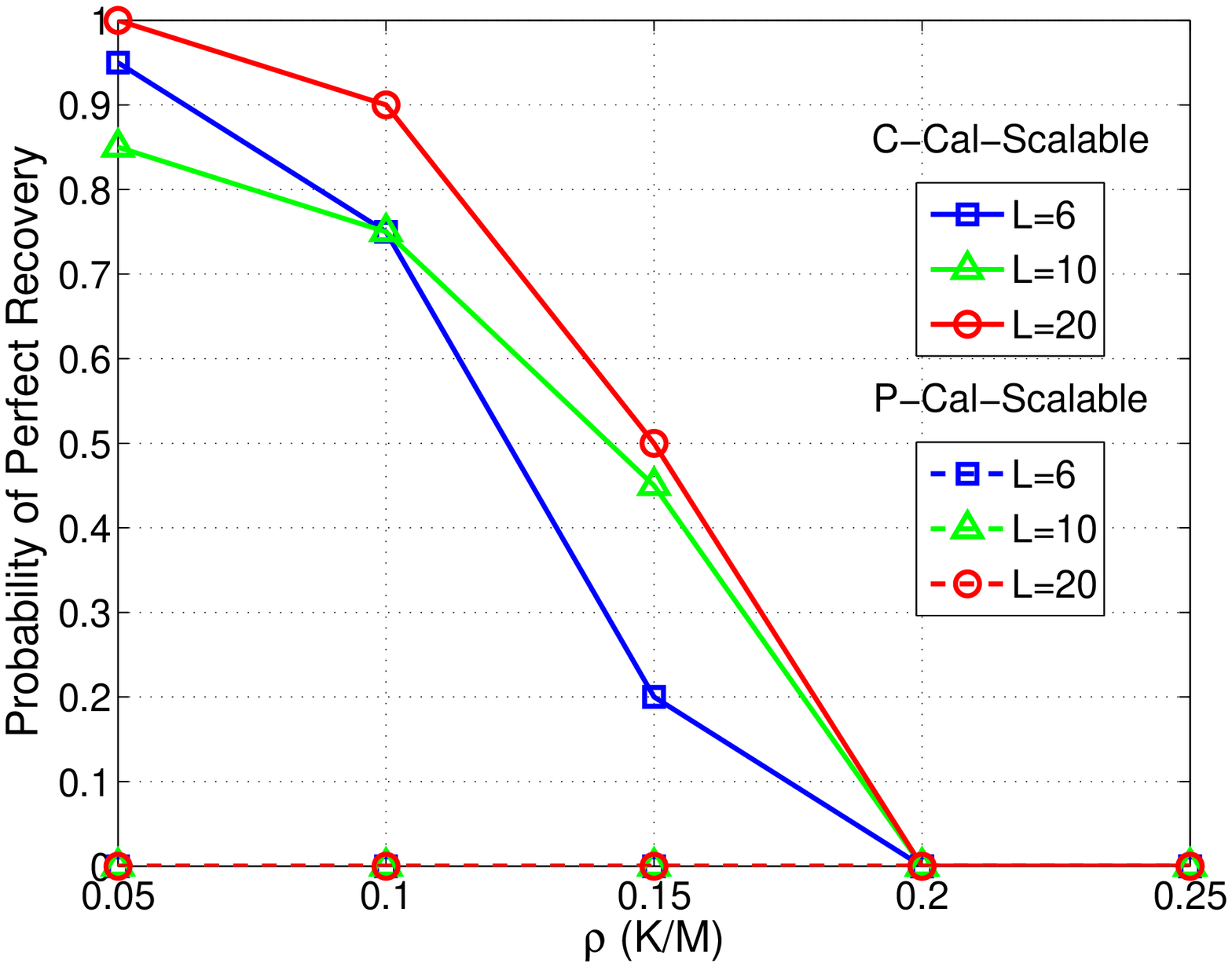}
}
\caption{({\CompCalS} and {\PhaseCalS}) The empirical probability of perfect recovery for $N=100$ and $\delta=0.8$ with respect to $\rho \triangleq K/M$. The algorithms {\CompCalS} (solid lines) and the {\PhaseCalS} (dashed lines) are compared for different amplitude variability, $\sigma$, and number of input signals, $L$.}
\label{fig:compcalex}
\end{figure*}

In this section, we consider the general gain calibration problem as described in Section~\ref{sec:PDef}, i.e. both the amplitude and the phases of the gains are varying among the sensors as shown in \eqref{eq:transfunc}. It has already been shown in Section~\ref{sec:AmpCal} that {\AmpCal} fails when the phase variability is very large. The optimization {\PhaseCalJ} (or {\PhaseCalS}) was developed assuming the sensors introduce only phase shifts, hence its performance can also be expected to be degraded heavily when the amplitude variability is large. As a result, none of these algorithms can be successful when there is considerable amplitude and phase variability.

In order to perform the complete calibration, it is possible to combine the amplitude and phase calibration approaches introduced in Sections~\ref{sec:AmpCal}, \ref{sec:PhaseCal} and \ref{sec:PhaseCalSc}. When there is unknown amplitudes of the gains, the cross measurements defined in \eqref{eq:crossmeas} are now equal to
\begin{align}
g_{i,k,\l} &\triangleq y_{i,k}y_{i,\l}\H\qquad  i=1\dots M,\: k,\l=1\dots L\\
&= d_i e^{j\theta_i} \m_i\H \xs_k \xs_\l\H \m_i e^{-j\theta_i} d_i\\
&= d_i^2\m_i\H \XS_{k,\l} \m_i\quad d_i \in \ReS
\end{align}
Since these measurements are scaled by non zero gains similar to the original measurements in \eqref{eq:transfunc}, it is possible to re-parameterize this equation in a linear form as in \eqref{eq:transfunc2} such that
\begin{align}
\label{eq:transfunc3}
\tau_ig_{i,k,\l} = \m_i\H \XS_{k,\l}\m_i \qquad &i=1,\dots, M\;,\; k,\l=1,\dots, L\\
\nonumber &\tau_i \triangleq \frac{1}{d_i^2}
\end{align}
Therefore we propose to address the complete calibration through the optimization
\begin{align}
\nonumber \CompCalJ\textbf{:} \qquad\qquad\qquad\quad &\\
\label{eq:compcal}
\begin{array}{c}
\Xh\\
\hat{\tau}_1,\ldots,\hat{\tau}_M
\end{array} = \argmin_{\Z,t_1,\ldots,t_M}\; &\Vert\Z\Vert_1 \\
\text{subject to}\; \nonumber & t_i g_{i,k,\l} = \m_i\H\Z_{k,\l}\m_i,\; i=1\dots M\\
\nonumber & \Z \succcurlyeq 0,  \qquad\qquad\quad\, k,\l=1\dots L\\
\nonumber &\sum_{n=1}^M t_n = c 
\end{align}
After the estimation of $\Xh$ and $\tau_i$  for $i=1,\ldots,M$, the unknown sparse signals are estimated from $\Xh$ as in {\PhaseCalJ}. The unknown gain amplitudes can be set as $\hat{d}_i = \frac{1}{\sqrt{\hat{\tau}_i}}$ and the gain phases $\hat{\theta}_i$ can be estimated trivially for all sensors using $\xh_\l$, $d_i$ and $y_{i,\l}$.

As seen in the performance of {\AmpCal} in Section~\ref{sec:AmpCalExp}, we can expect the minimum number of input signals for perfect recovery to grow as the amplitude variability increases in the complete calibration. Therefore the size of the optimization problem in {\CompCalJ} can increase greatly due to dealing with a number of unknowns growing as $L^2$. To better handle this issue, a scalable version of {\CompCalJ} can be formulated as
\begin{align}
\nonumber {\CompCalS}\textbf{:}\qquad\quad & \\
\label{eq:compcal_fast}
\nonumber \begin{array}{c}
\Xh_{1,1},\ldots,\Xh_{L,L}\\
\Xh_{1,2},\ldots,\Xh_{L,1}\\ 
\hat{\tau}_1,\ldots,\hat{\tau}_M
\end{array} =\quad & \\
\argmin_{\substack{\Z_{1,1}\dots\Z_{L,L}\\ \Z_{1,2}\dots\Z_{L,1}\\t_1,\ldots,t_M}} \sum\limits_{k=1}^L \left( \Vert\Z_{k,k}\Vert_1 \right. & + \left. \Vert\Z_{k,(k)_L+1}\Vert_1 \right) \\
\nonumber \text{subject to} \qquad t_i g_{i,\l,\l} = & \m_i\H\Z_{\l,\l}\m_i \qquad\qquad\: i=1,\dots, M\\
\nonumber t_i g_{i,\l,(\l)_L+1} = & \m_i\H\Z_{\l,(\l)_L+1}\m_i \qquad \l=1,\dots, L  \\
\nonumber \left[ \begin{array}{l}
\Z_{\l,\l}\\
\Z_{\l,(\l)_L+1}\H
\end{array} \right. & \left. \begin{array}{l}
\Z_{\l,(\l)_L+1}\\
\Z_{(\l)_L+1,(\l)_L+1}
\end{array} \right] \succcurlyeq 0\\
\nonumber \sum_{n=1}^M t_n &= c
\end{align}
which is simply the combination of the optimization approaches in {\AmpCal} and {\PhaseCalS}. As a result of the optimization, the estimated signals, $\xh_\l$, are set as in {\PhaseCalS} whereas the estimated gains $\hat{d}_i$ and the phases $\hat{\theta}_i$ can be set as in {\CompCalJ}.

\subsection{Experimental Results for Complete Calibration}

When there is no amplitude variability, the performance of {\CompCalJ} can be expected to be similar to the performance of {\PhaseCalJ} considering the performance of {\AmpCal} and {\PhaseCalJ} in Sections~\ref{sec:AmpCal}-\ref{sec:PhaseCalSc}. Similar to the performance of {\AmpCal}, as the amplitude variability increases, this performance is expected to worsen unless sufficiently many input signals are jointly recovered. Since the number of input signals required for best performance grows with to the amplitude variability, the computational scalability in $L$ is of prime importance for the complete calibration scenario. 

Considering the higher complexity of the problem and the sheer number of cases to be simulated, we shall not provide complete phase transition diagrams for complete calibration as in the earlier sections. Instead, we demonstrate the performances of {\PhaseCalS} and {\CompCalS} (both implemented with ADMM) in the presence of amplitude variability for a specific number of measurements ($\delta=\frac{M}{N}=0.8$). {\PhaseCalJ} and {\CompCalJ} are not simulated due to the intractable complexity of the simulations for large $L$. Figure~\ref{fig:compcalex} shows the comparison of the two methods in terms of the empirical probability of perfect recovery as a function of $\rho=\frac{K}{M}$ under different levels of amplitude variability ($\sigma=0.1, 0.3, 0.5$) and number of input signals ($L=6, 10, 20$). The experimental setup and the criteria for perfect recovery are identical to the experiments in Sections~\ref{sec:PhaseCal} and \ref{sec:PhaseCalSc} while the probability of perfect recovery is empirically estimated over 20 independent simulations. The method {\AmpCal} is not compared in the experiments since it is shown to fail under maximum phase variability in Section~\ref{sec:AmpCal}. The results in Figure~\ref{fig:compcalex} demonstrate that when there is large amplitude and phase variability in the gains, phase calibration approaches presented earlier has no hope of success, whereas the complete calibration method {\CompCalS} can perform blind calibration successfully. However it can also be noticed that the contribution of additional number of input signals is much smaller compared to what has been observed with the method {\AmpCal}. 

\section{Conclusions}
\label{sec:Conc}

\begin{figure*}[t]
\centering
\includegraphics[width=.65\textwidth]{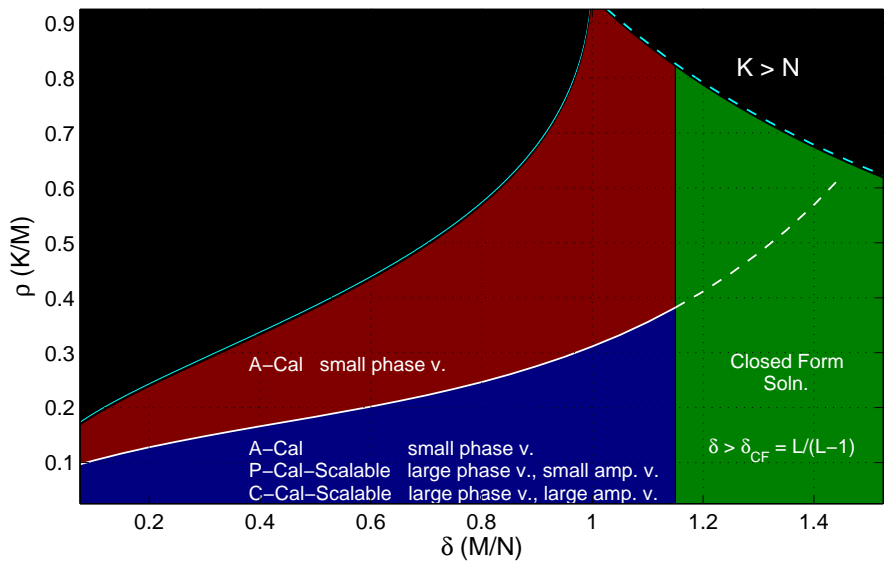}
\caption{A depiction of the algorithms to be used for different calibration scenarios. Note that for each case $L$ must also be sufficiently high to guarantee successful recovery depending on the amplitude and phase variability. The solid cyan line indicates the Donoho-Tanner phase transition curve for fully calibrated compressed sensing recovery \cite{Donoho2009}.}
\label{fig:summary}
\end{figure*}

Several convex optimization methods to handle blind calibration of complex-valued gains in sparse inverse problems have been proposed in this paper\footnote{The codes for the MATLAB\textsuperscript{\textregistered} implementations of the proposed methods has been provided in\\ http://hal.inria.fr/docs/00/96/02/72/TEX/Calcodesv2.0.rar}. Upon investigating the performance of all of the presented methods, it can be seen that each one provides complementary advantages and can be used in a different scenario depending on the distribution of the unknown sensor gains. 

A summary of which approach to use for different conditions on the sensor gains is shown in Figure~\ref{fig:summary} along with the successful recovery regions of the algorithms in the $\delta-\rho$ plane. 
\begin{itemize}
\item When $\delta \geq \dcf$ (indicated with green color in the figure), a unique solution satisfying the measurements can be found without any optimization, regardless of the distribution of the gains and the input signals.
\item When $\delta < \dcf$ and phase variability is not large, {\AmpCal} is preferable over the other methods due to its significantly lower complexity and higher performance. Provided that there are sufficiently many input signals, {\AmpCal} can achieve recovery performance similar to calibrated recovery in \eqref{eq:CalibratedL1} (indicated with the regions colored in red and blue in the figure).
\item  When the phase variability is large, {\CompCalS} or {\PhaseCalS} can be used depending on whether there is any amplitude variability or not respectively. These quadratic methods can perform successful recovery when the input signals are sufficiently sparse (indicated with the blue region in the figure).
\end{itemize}  

We have also shown that the scalable quadratic methods ({\PhaseCalS} and {\CompCalS}) are preferable to their non-scalable counterparts ({\PhaseCalJ} and {\CompCalJ}) since they provide significant reduction in optimization complexity with negligible performance loss compared to the latter. Unfortunately, none of the proposed methods can achieve recovery for low sparsity signals (indicated with red region in the figure) when the phase variability is very large. Handling this case is considered as future work. A detailed theoretical analysis of the proposed algorithms in terms of sample complexity and performance bounds is the focus of future study as well.


The proposed methods in this paper have been evaluated under the assumption of a noiseless system and the performance in the presence of additive noise is beyond the scope of this paper. The initial impression from the linearized constraint equations (\eqref{eq:transfunc2} and \eqref{eq:transfunc3}) can be that dealing with the inverse of the gains could affect the performance under noise due to the noise becoming multiplicative instead of additive. Our initial empirical results rather suggest that the recovery performance is not severely affected unless the amplitude variability is very large, which applies to many practical applications. Further investigation of the performance of the proposed methods in the presence of noise is also considered as future work.



\bibliographystyle{IEEEtran}
\bibliography{calibration.bib}

\vfill\eject

\begin{IEEEbiography}[{\includegraphics[width=1in,height=1.25in,clip,keepaspectratio]{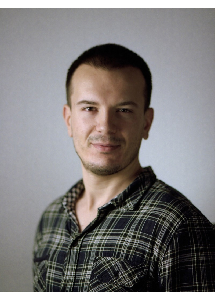}}]{\c{C}a\u{g}da\c{s}~Bilen}
\c{C}a\u{g}da\c{s}~Bilen received his BS degree in Electrical and Electronics Engineering in 2005 from Middle East Technical University (METU), Ankara, Turkey. He continued his graduate studies in METU under the supervision of Gozde Bozdagi Akar and received his MS degree in 2007. He received his PhD. degree in Electrical Engineering in Polytechnic Institute of New York University in 2012 under the co-supervision of Prof. Yao Wang and Prof. Ivan W. Selesnick. 

Since 2012, he has been a post-doctoral researcher in PANAMA team led by Remi Gribonval in INRIA Rennes, France (French Institute for Research in Computer Science and Control). His research interests include linear inverse problems, sparse recovery, compressed sensing, blind calibration, magnetic resonance imaging, video and image compression, 3-D and multiview video processing. 
\end{IEEEbiography}

\begin{IEEEbiography}[{\includegraphics[width=1in,height=1.25in,clip,keepaspectratio]{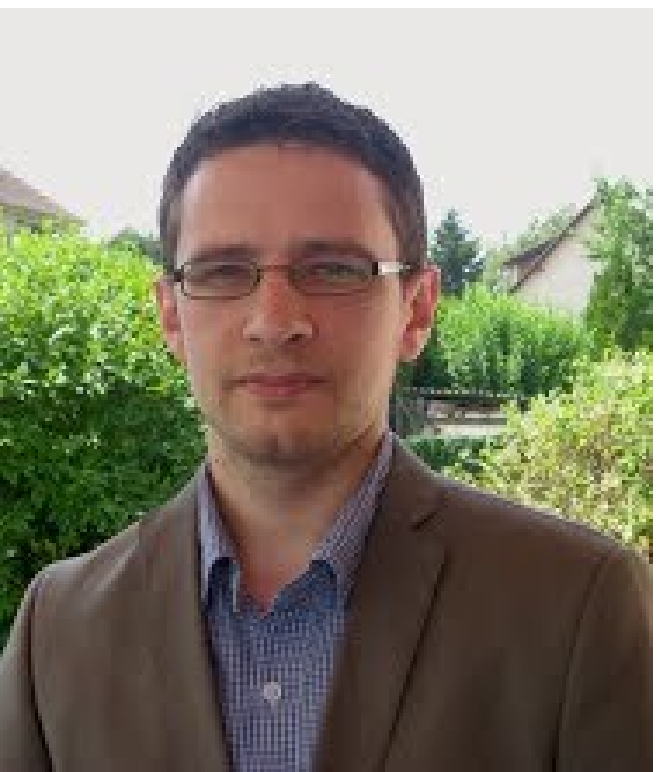}}]{Gilles Puy}
Gilles Puy received the M.Sc. degree in Electrical and Electronics Engineering from Ecole Polytechnique Fédérale de Lausanne (EPFL), Switzerland, and the Engineering Diploma (M.Sc.) from Supélec, France, in 2009. He received the Ph.D. degree in Electrical Engineering from EPFL in January 2014. His Ph.D. advisor was Prof. Pierre Vandergheynst. In 2013, he was awarded the EPFL Chorafas Foundation award for his Ph.D. work.

He is currently a postdoctoral researcher at INRIA, France, where he is a member of the PANAMA Research team, led by Dr. Rémi Gribonval. His research focuses mainly on compressive sampling, inverse problems, convex and non-convex optimisation.
\end{IEEEbiography}

\begin{IEEEbiography}[{\includegraphics[width=1in,height=1.25in,clip,keepaspectratio]{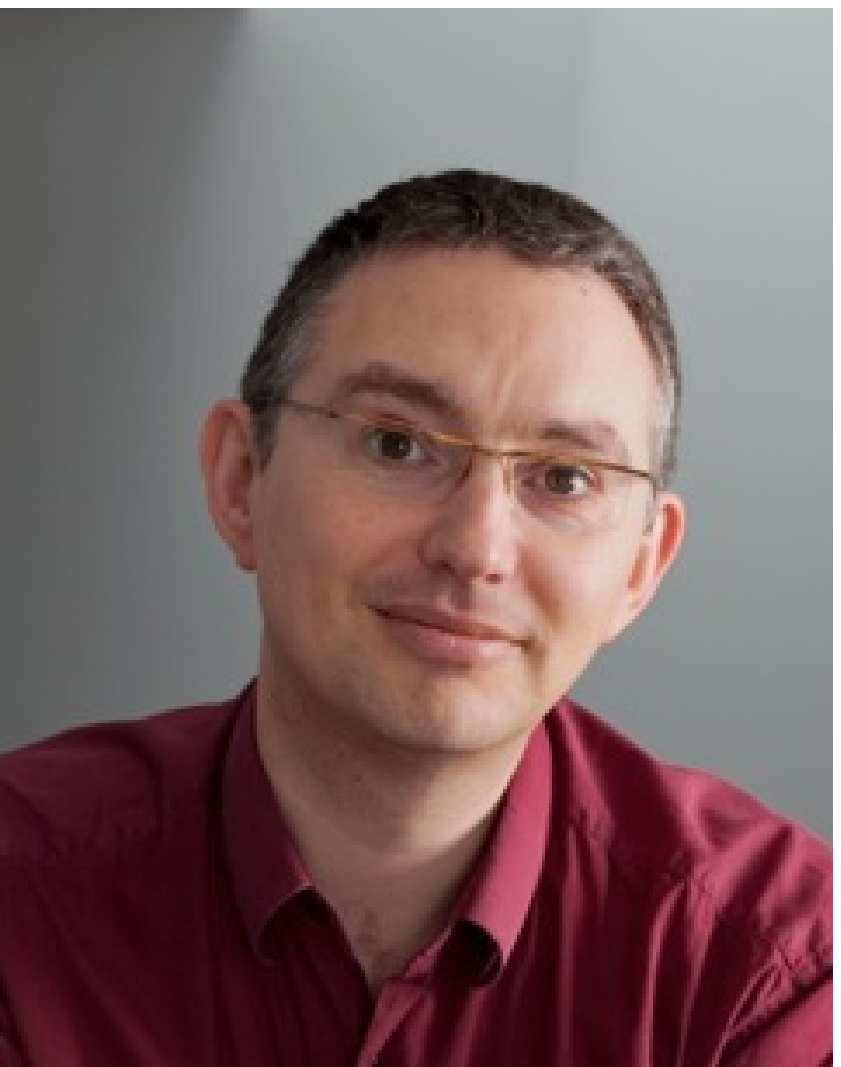}}]{R\'emi Gribonval}
R{\'e}mi Gribonval(FM'14)  is a Senior Researcher with Inria (Rennes, France), and the scientific leader of the PANAMA research group on sparse audio processing. A former student at  {\'E}cole Normale Sup{\'e}rieure (Paris, France), he received the Ph. D. degree in applied mathematics from Universit{\'e} de Paris-IX Dauphine (Paris, France) in 1999, and his Habilitation {\`a} Diriger des Recherches in applied mathematics from Universit{\'e} de Rennes~I (Rennes, France) in 2007. His research focuses on mathematical signal processing, machine learning, approximation theory and statistics, with an emphasis on sparse approximation, audio source separation, dictionary learning and compressed sensing. 
He founded the series of international workshops SPARS on Signal Processing with Adaptive/Sparse Representations. In 2011, he was awarded the Blaise Pascal Award in Applied Mathematics and Scientific Engineering from the SMAI by the French National Academy of Sciences, and a starting investigator grant from the European Research Council. 
\end{IEEEbiography}

\begin{IEEEbiography}[{\includegraphics[width=1in,height=1.25in,clip,keepaspectratio]{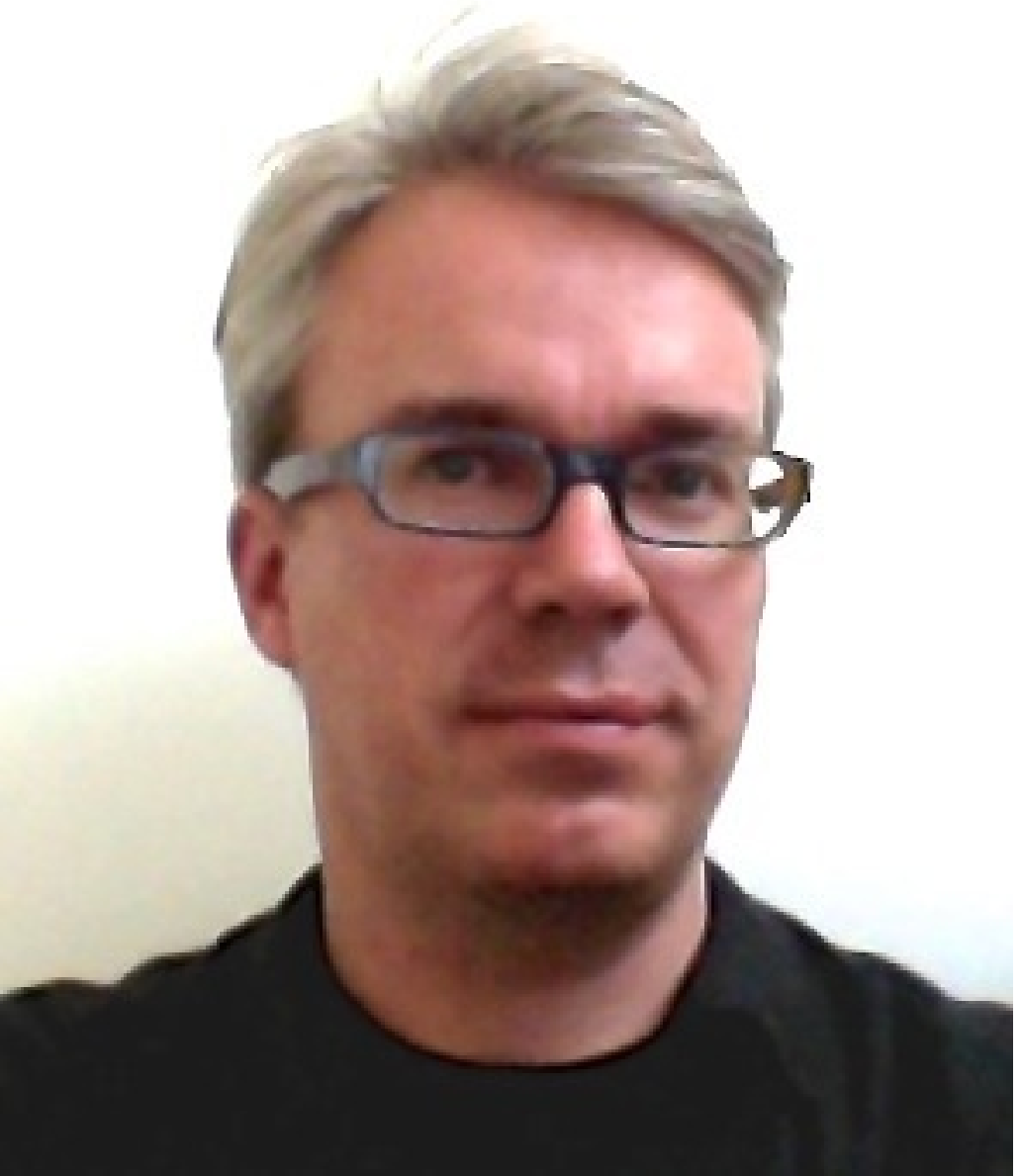}}]{Laurent Daudet}
Laurent Daudet (M’04–SM’10) studied at the Ecole Normale Supérieure in Paris, where he graduated in statistical and non-linear physics. In 2000, he received a PhD in mathematical modeling from the Université de Provence, Marseille, France. After a Marie Curie post-doctoral fellowship at the C4DM, Queen Mary University of London, UK, he worked as associate professor at UPMC (Paris 6 University) in the Musical Acoustics Lab. He is now Professor at Paris Diderot University – Paris 7, with research at the Langevin Institute for Waves and Images, where he currently holds a joint position with the Institut Universitaire de France. Laurent Daudet is author or co-author of over 150 publications (journal papers or conference proceedings) on various aspects of acoustics and audio signal processing, in particular using sparse representations. 
\end{IEEEbiography}

\end{document}